\pdfoutput=1
\documentclass[12pt,letterpaper]{article}
\usepackage{jheppub}

\usepackage{latexsym}
\usepackage{cancel}
\usepackage{subfigure,bbm}
\usepackage{graphicx}
\usepackage{amssymb,amsmath,psfrag,epsfig}

\newcommand{\be}{\begin{eqnarray}}
\newcommand{\ee}{\end{eqnarray}}
\newcommand{\beq}{\begin{equation}}
\newcommand{\eeq}{\end{equation}}
\newcommand{\beqa}{\begin{eqnarray}}
\newcommand{\eeqa}{\end{eqnarray}}
\newcommand{\bea}{\begin{aligned}}
\newcommand{\eea}{\end{aligned}}
\newcommand{\bem}{\begin{multline}}
\newcommand{\eem}{\end{multline}}
\newcommand{\nn}{\nonumber  \\}

\newcommand{\Gd}{\delta}

\newcommand{\Gl}{\lambda}

\def\smcon#1{\sum_{i=1}^{#1}\lambda_i \eta_i}

\newcommand{\A}{\mathbb{A}}

\def\bra(#1,#2){[#1\,#2]}
\def\ket(#1,#2){\langle #1\,#2\rangle}
\def\mt(#1,#2,#3,#4){\langle #1\,#2\,#3\,#4\rangle}

\def\mt(#1){\mathcal{Z}_{#1}}
\newcommand{\ab}[1]{\langle #1\rangle}

\def\m#1{#1^-}

\def\N#1{\mathcal{N}=#1}

\title{\textbf{Generating All Tree Amplitudes in $\N4$ SYM by Inverse Soft Limit}}

\author[a]{Dhritiman Nandan}
\author[b]{and Congkao Wen.}

\affiliation[a]{Department of Physics, Brown University, Box 1843, Providence, RI 02912-1843, USA}
\affiliation[b]{Centre for Research in String Theory, School of Physics and Astronomy, Queen Mary University of London, Mile End Road, London, E1 4NS, United Kingdom}

\emailAdd{dhritiman\_nandan@brown.edu}
\emailAdd{c.wen@qmul.ac.uk}

\abstract{The idea of adding particles to construct amplitudes has been utilized in various ways in exploring the structure of scattering amplitudes.  This idea is often called Inverse Soft Limit, namely it is the reverse mechanism of taking particles to be soft. We apply the Inverse Soft Limit to the tree-level amplitudes in $\mathcal{N}=4$ super Yang-Mills theory, which allows us to generate full tree-level superamplitudes by adding ``soft" particles in a certain way. With the help from Britto-Cachazo-Feng-Witten recursion relations, a systematic and concrete way of adding particles is determined recursively. The amplitudes constructed solely by adding particles not only have manifest Yangian symmetry, but also make the soft limit transparent. The method of generating amplitudes by Inverse Soft Limit can also be generalized for constructing form factors.}

\begin{document}
\maketitle

\section{Introduction}

There have been enormous progress in unraveling the deep structure of scattering amplitudes in gauge theories and gravity. Those beautiful mathematic properties of the scattering amplitudes are often hidden in the complexity of their traditional Feynman diagram expansion. In the past years various powerful tools beyond Feynman diagrams have been largely developed, such as (generalized) unitarity-cut method~\cite{generalizedunitarity,Britto:2004nc}, and Britto-Cachazo-Feng-Witten (BCFW) recursion relations and MHV rules at both tree level~\cite{BCFW,ArkaniHamed:2008yf,Cachazo:2004kj} and loop level~\cite{ArkaniHamed:2010kv,Brandhuber:2004yw}, and more recently a new and exciting mathematical tool called Symbols has been proved to be greatly useful in the calculation of loop amplitudes~\cite{Goncharov:2010jf}. All of those tools (as well as many not mentioned others)  not only allow us to carry out previous impossible calculations efficiently, but also to discover beautiful, but hidden structures of the theories. Indeed it were only recent years, many unexpected structure of scattering amplitudes were discovered, which include the dual conformal symmetry and Yangian symmetry~\cite{Drummond:2008vq} along with the duality between scattering amplitudes and Wilson loop (correlation functions)~\cite{Wilsonloop,correlation}  in $\mathcal{N}=4$ Super Yang-Mills (SYM) theory, and possible ultraviolet finiteness in $\mathcal{N}=8$ super gravity~\cite{gravity}, as well as newly discovered duality between color and kinematics~\cite{Bern:2008qj}.

It has been known for a long time that, for gauge theories and gravity, under the soft limit scattering amplitudes of any number of external particles reduces to amplitudes with one less number of external particle times an universal soft factor\cite{Weinberg}. It is an amazing fact that amplitudes of gauge theories and gravity behave nicely under the soft limit. The study of soft limit of scattering amplitudes in field theories has been remarkably successful in understanding their structure. In fact, soft limit (as well as collinear limit) have been extensively used as strong constraints for helping to fix the scattering amplitudes~\cite{soft}.

In recent years there had been a surging interest in understanding how to build up amplitudes by adding particles starting from an amplitude with lower number of particles, which is exactly the reverse mechanism of taking a soft limit. So, under this paradigm, also called the ``Inverse Soft", the soft behaviour of scattering amplitudes are just enough to restrict the structure of amplitudes. This phenomenon was first observed and suggested in~\cite{ArkaniHamed:2009talk,ArkaniHamed:2009si}, where the scattering amplitude was described in terms of the Hodges' diagram representation~\cite{Hodges:2005aj}, and later was introduced as one of the important ingredients in the Grassmannian approach to the scattering amplitudes~\cite{ArkaniHamed:2009sx,ArkaniHamed:2009dg,ArkaniHamed:2009dn,ArkaniHamed:2010kv,Dolan:2009wf,Spradlin:2009qr,Nandan:2009cc,Bourjaily:2010kw,Mason:2009qx,Bullimore:2009cb}. For the applications of Inverse Soft Limit (ISL) in understanding various aspects of the scattering amplitudes in $\mathcal{N}=4$ Super Yang-Mills (SYM) theory and $\mathcal{N}=8$ super gravity as well, see for instance~\cite{Nguyen:2009jk,Bullimore:2010pa,Dunbar:2011dw,Dunbar:2012aj,Hodges:2011wm,Hodges:2012ym}. 

Moreover, it is known that from the point of view of the Grassmannian, ISL~\cite{ArkaniHamed:2009dg,ArkaniHamed:2010kv} is a natural way of constructing Yangian-invariants~\cite{Drummond:2008vq}. Since tree-level amplitudes in $\N4$ SYM are Yangian invariant we should be able to construct these amplitudes by the above mentioned ISL, but unfortunately a systematic way of doing this had so far eluded us. In this paper we address this issue. In a recent paper~\cite{BoucherVeronneau:2011nm} some progress had been made in carrying out this program for few simple non-supersymmetric amplitudes in $\N4 $ SYM. This process is related to the two particle factorization channel of BCFW recursion relations. In this paper we generalize their results and show that all superamplitudes in $\N4 $ SYM at tree level can be constructed by an explicit prescription of ISL, namely by a systematic way of adding a series of particles to lower-point superamplitudes to arrive at higher-point superamplitudes. By analysing and examining the BCFW diagrams carefully, we are able to obtain recursion relations of how to construct an arbitrary BCFW diagram by adding particles from any side of that diagram, consequently any arbitrary amplitude will be ISL constructible in a concrete way. It is clear that the amplitudes constructed solely by adding particles not only have manifest Yangian symmetry, but also make the soft limit transparent. 

The ISL in $\N4$ SYM is closely tied to the ideas of Grassmannian formalism, Hodges' diagrams, as well as Yangian invariance, and indeed the amplitudes constructed by ISL are guaranteed to have Yangian symmetry, as we mentioned. However the canonical configuration for adding particles obtained from the proposed recursion relations is quite independent of those ideas. In fact our way of adding particles can be straightforwardly generalized to another interesting class of physical observables, the form factors~\cite{vanNeerven:1985ja}, for which all of above mentioned ideas may fail. So the ISL prescription constructed in this paper is quite general. On the other hand, the fact that ISL can be applied to the form factors may indicate that there might also exist hidden symmetries in form factors. In fact, indeed the tree-level solutions for form factors resemble the tree-level solutions for $\mathcal{N}=4$ SYM theory~\cite{DH08}, which is known to be manifestly dual conformal invariant.\footnote{We would like to thank Gang Yang for the discussion on this topic.}

The paper is organized as follows. In Section $2$ we review the idea that two particle factorization channel of BCFW recursion relations is related to the notion of adding a particle to a tree amplitude in $\N4$ SYM by ISL. We also present a few non-trivial examples of constructing amplitudes using the ISL method. Then we move on to section $3$ where we determine the canonical configuration for adding particles to construct BCFW terms, and consequently a concrete prescription to generate any tree level superamplitudes of $\N4$ by this method is given. This canonical configuration takes the form of a set of recursion relations where we can generate a higher point configuration from lower point ones. In section $4$, the BCFW shifts of a BCFW diagram have been shown to be in one-to-one correspondence with certain multiple shifts in the ISL picture. We go on to extend the ISL paradigm to construct form factors of $\N4$ SYM in section $5$. Example using our recursion relation and discussions on the extension of the ISL method to gravity amplitudes and ISL in the momentum-twistor language are presented in the Appendix.

\section{Two-particle channel BCFW and ISL in SYM}
\subsection{Inverse soft factors and shifts}
In~\cite{BoucherVeronneau:2011nm} it was shown that the notion of adding a particle to a tree-level amplitude by ISL is related to the two particle factorization channel of BCFW recursion relations. In this section we review and extend their results for the supersymmetric BCFW recursion relations. Before we proceed let us mention that  we would be using the symbol $\langle 1~n]$ to denote the following BCFW shifts
\be \label{bcfwshift}
\lambda_{\hat{1}} &=& \lambda_1 -z \lambda_n; \\ \nonumber
\tilde{\lambda}_{\bar{n}} &=& \tilde{\lambda}_{n} + z \tilde{\lambda}_{1}; \\ \nonumber
\eta_{\bar{n}} &=& \eta_n + z \eta_1, 
\ee
and $[1~n \rangle $ to denote the parity flipped version of the above shifts, namely
\be \label{Pbcfwshift}
\lambda_{\bar{1}} &=& \tilde{\lambda_1} -z \tilde{\lambda_n}; \\ \nonumber
{\lambda}_{\hat{n}} &=& {\lambda}_{n} + z {\lambda}_{1}; \\ \nonumber
\eta_{\bar{1}} &=& \eta_1 - z \eta_n. 
\ee
We note here that we encounter two different BCFW diagrams for two particle factorization channel, and we will soon explain that they correspond to the cases where we add a positive and a negative helicity particles to a lower-point amplitude in the non-supersymmetric case, which are respectively called $k$ preserving and $k$ increasing inverse-soft operations in the supersymmetric case (see the discussion in Appendix A.), here $k$ denotes the degree of R-charges of N$^k$MHV\footnote{The meaning of this notation will become clear shortly.} amplitudes. 

\begin{figure}[t]
\begin{center} 
\includegraphics[width=6cm]{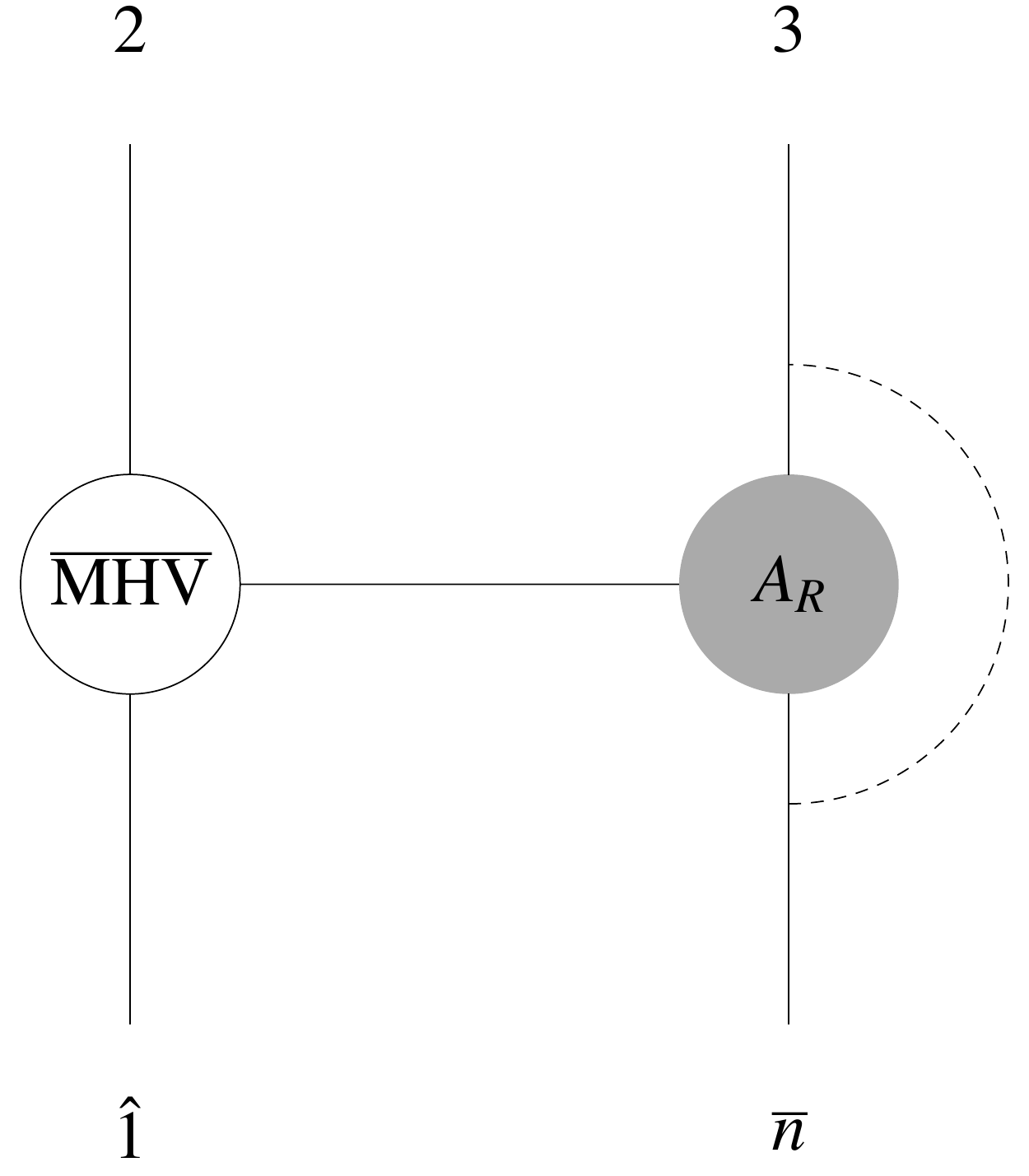}
\caption{\small{BCFW diagram of two particle channel corresponding to adding particle $1^+$.}} 
\label{BCFW1hat}
\end{center}
\end{figure}

Let us start the discussion with $\langle 1~n]$ BCFW shifts, see Fig.(\ref{BCFW1hat}), we have
\be
A(\hat{1},2|3,\cdots,\bar{n})=\int d^4 \eta_{\hat{P}}  A_L(\hat{1}, 2, -\hat{P}) {1 \over s_{12}} A_R(\hat{P}, 3, \dots, \bar{n}).
\ee
\!\!Note that because of the particular choice of the BCFW shift, the three-point amplitude $A_L(\hat{1}, 2, -\hat{P})$ must be a $\overline{\rm MHV}$ amplitude, namely the parity flipped version of the maximally-helicity-violating (MHV) amplitudes. It is straightforward to find that this BCFW diagram can be written as,
\be
A(\hat{1},2|3,\cdots,\bar{n})=\mathcal{S}_+(n~1~2) A_R(2', 3, \dots , n' ), 
\ee
where the soft factor $\mathcal{S}_+(n~1~2)$ is defined as,
\be 
\mathcal{S}_+(n~1~2) ={\langle n 2 \rangle \over \langle n 1\rangle \langle 1 2 \rangle}.
\label{Splus}
\ee
Here the primed particle labels, $2'$ and $n'$, represent the following shifts on particles $2$ and $n$,
\be \label{shift}
\tilde{\lambda}_2 \rightarrow  \tilde{\lambda}_2 + {\langle 1 n \rangle \over \langle 2 n \rangle} \tilde{\lambda}_1, \, \tilde{\lambda}_n \rightarrow  \tilde{\lambda}_n + {\langle 1 2 \rangle \over \langle n 2 \rangle} \tilde{\lambda}_1 ;  \\ \nonumber
\eta_2 \rightarrow  \eta_2 + {\langle 1 n \rangle \over \langle 2 n \rangle} \eta_1, \, \eta_n \rightarrow  \eta_n + {\langle 1 2 \rangle \over \langle n 2 \rangle} \eta_1.
\ee 
The shifts ensure the momenta and supercharge conservation after adding particles. As indicated in $\mathcal{S}_+(n~1~2)$, this case will often be called as adding a positive particle $1^+$, although we are dealing with a supersymmetric amplitude.
\begin{figure}[t]
\begin{center} 
\includegraphics[width=6cm]{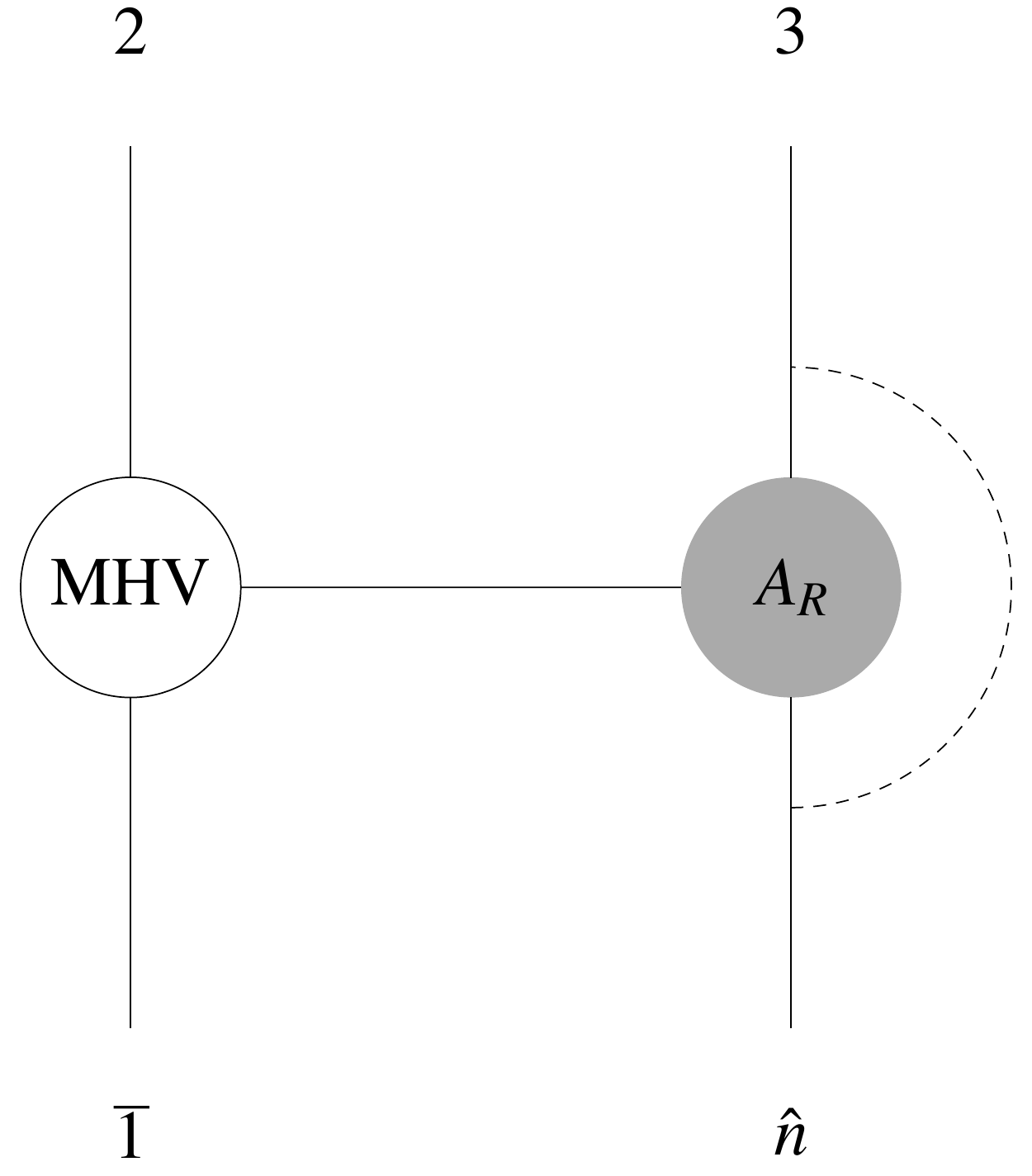}
\caption{\small{BCFW diagram of two particle channel corresponding to adding particle $1^-$.}} 
\label{BCFW1hatn}\end{center}
 \end{figure}
 
A similar calculation for the parity flipped version of the previous case i.e. for ~(\ref{Pbcfwshift}), see Fig.(\ref{BCFW1hatn}), leads to another kind of soft factor, which is given as
\beqa
\mathcal{S}_-(n~1~2) &=& {[n2] \over [n 1][12]}  \delta^4(\eta_1 + {[n1] \over [2n]} \eta_2 + {[12] \over [2n]} \eta_{n})\nn
&=& {1\over [n1][12][n2]^3}\delta^4([12]\eta_n + [2n]\eta_1 + [n1]\eta_{2}),
\label{Sminus}
\eeqa
with the following ISL shifts
\be \label {shift2}
{\lambda}_n \rightarrow  {\lambda}_n + {[21] \over [2n]} {\lambda}_1, \, 
{\lambda}_{2} \rightarrow  {\lambda}_{2} + {[1n] \over [2n]}{\lambda}_1. 
\ee
We note that ${[n2] \over [n 1][12]}$ is the soft factor for removing a negative particle, while the extra fermionic delta function takes care of increasing the R-charge. Moreover the higher-point superamplitude we get by adding particle $1^-$ is given by
\beq
A(\bar{1},2|3,\ldots, \hat{n})=\mathcal{S}_-(n~1~2) A(2', \ldots, (n-1),n'),
\eeq
again $2'$ and $n'$ indicate the shifts on $2$ and $n$ according to \eqref{shift2}. 

Let us conclude this subsection with remarks on how to generate general BCFW diagrams with multiple-particle channels according to ISL. It is clear that the previous discussion only allows us to rewrite BCFW diagrams with two-particle channel in the ISL form. To deal with a BCFW diagram with a multiple-particle channel, some attempt has been made in~\cite{BoucherVeronneau:2011nm}, where the goal is to build up a general BCFW diagram by adding particles to a two-particle-channel BCFW diagram. For instance let us consider a typical BCFW diagram for the $[1~n\rangle$ shift,
\beqa
A_L(\bar{1}, 2, \cdots, m,\hat{P}){1\over P^2} A_R(-\hat{P}, m\!+\!1, \cdots, n\!-\!1,\hat{n}).
\eeqa
The idea is to start with a two-particle-channel diagram, which we know how to write in the ISL form,
\beq
 A_L(\bar{1}, m,\hat{P}){1\over P^2} A_R(-\hat{P}, m\!+\!1, \cdots, n\!-\!1,\hat{n}) 
= \mathcal{S}(n~1~m) A_R(m', m\!+\!1, \cdots, n\!-\!1, n').
\eeq
We then build up the full subamplitude $A_L(\bar{1}, 2, \cdots, m,\hat{P})$ gradually by adding particles between $1$ and $m$. \textit{A priori} it is not guaranteed that $A_L(\bar{1}, 2, \cdots, m,\hat{P})$ can be constructed in this way. And indeed in the non-supersymmetric case, it was checked in~\cite{BoucherVeronneau:2011nm} that only few simple amplitudes can be constructed in such a way. From our previous discussion on the relation between ISL and the two-particle-channel BCFW diagram, we have seen that it is very natural to consider ISL for the superamplitudes in $\N4$ SYM theory. In fact, supersymmetry provides a huge advantage as we will show in the following sections that one can actually construct full superamplitudes in $\N4$ SYM solely by adding particles according to ISL.  Let us state our final result here before we proceed further.

We can proceed in the above mentioned way and generate any tree-level super amplitude in $\N4$ SYM theory by ISL and this can be schematically written as, 
\be \label{ISLampinitial}
A_n =\sum_{i;L,R} (\prod_L \mathcal{S}'_L) (\prod_R \mathcal{S}'_R) A_{\overline{\rm MHV}}(i',i\!+\!1,n'),
\ee
where summation over $i$ is according to BCFW diagrammatic representation of the amplitudes. The products on $\mathcal{S}_L$ and $\mathcal{S}_R$ and summation on $L, R$ are determined by a set recursion relations (\ref{recursionNkbar}) and (\ref{recursionNkhat}) which we propose in the subsequent sections to generate the configurations of particles that had to be added on both sides of a BCFW diagram to generate the BCFW diagram. And finally $\mathcal{S}'$, $i'$ and $n'$ are used for the fact that the particles are shifted according to the rules of ISL, namely Eq.~(\ref{shift}) and (\ref{shift2}).

\subsection{Examples}

Before we consider the general procedure for adding particles, let us consider a couple of simple examples to illustrate the idea of ISL. The first case we would like to consider is $\overline{\rm MHV}$ amplitude. As per the general philosophy we will start with a three-point amplitude and gradually build up the full amplitude by adding particles. Let us consider five-point amplitude first, one way of doing this is following: 
 \begin{itemize}
 \item We start by adding the particle $1^+$ to $A_{\rm MHV}(345)$ in order to generate $A_{\rm MHV}(1345)= \mathcal{S}_+(513)A_{\rm MHV}(3'45')$, where the shifts are according to \eqref{shift}. To show this we note that,
 \beqa
 A_{\rm MHV}(1345)&=& \frac{\ket(5,3)}{\ket(5,1)\ket(1,3)}\bigl( \frac{\Gd^8(\Gl_{3'}\eta_{3'}+\Gl_4\eta_4+\Gl_{5'}\eta_{5'})}{\ket(3',4)\ket(4,5')\ket(5',3')}\bigr)\nn
 &=& \frac{\Gd^8(\Gl_1\eta_1+\Gl_{3}\eta_{3}+\Gl_4\eta_4+\Gl_{5}\eta_{5})}{\ket(1,3)\ket(3,4)\ket(4,5)\ket(5,1)},
 \eeqa
 \textit{where we simplified the supercharge conserving delta function by Scouten identities.}
 \item Our final goal $A_{\rm NMHV}(12345)$ can be obtained by further adding  $2^-$ to $A_{\rm MHV}(1345)$,
\beqa
A_{\overline{\rm MHV}}(12345)&=&\mathcal{S}_-(123) A_{\rm MHV}(1'3'45)\nn
&=&\frac{\Gd^4(\eta_1[23] +\eta_2[31]+\eta_3[12])\Gd^8(\smcon5)}{\bra(1,2)\bra(2,3)\bra(3,4)\bra(4,5)\bra(5,1)\ket(4,5)^4},
\label{a5nmhv}
\eeqa
where $1'~{\rm and}~ 3'$ are shifted according to (\ref{shift2}). \textit{We simplified the supercharge conserving delta function by using the fermionic delta function from $S_-$.}
\end{itemize} 

One can continue the process and add a negative helicity particle to \eqref{a5nmhv}. One particular way we are using here, as the five-point case, is to add $2^-$ between $1$ and $3$ to the five-point amplitude $A_{\overline{\rm MHV}}(13456)$ and we get,\footnote{One of course could add particles in a different way, the answer would be in a different-looking form.}
\beq
A_{\overline{{\rm MHV}}}(123456)=\frac{\Gd^8(\sum \Gl_i \eta_i)\Gd^4(\eta_1[34]+\eta_3[41]+\eta_4[13])\Gd^4(\eta_1[23]+\eta_2[31]+\eta_3[12])}{([12][23]\ldots [61])\bra(1,3)^4\ket(5,6)^4}.
\eeq
Similarly a compact general formula for $n$-point $\overline{\rm MHV}$ amplitudes can be obtained by continuing to add $2^-$ between $1~{\rm and}~3$,
\beq
A_{\overline{{\rm MHV}}}(1,2,\cdots,n)=\frac{\Gd^8(\sum \Gl_i \eta_i)\prod_{i=2}^{n-3}\Gd^4(\eta_1\bra(i,i+1)+\eta_i\bra(i+1,1)+\eta_{i+1}\bra(1,i))}{\ket(n\!-\!1,n)^4\prod_{i=1}^{n}\bra(i,i+1)\prod_{i=3}^{n-3}\bra(1,i)^4}.
\label{Amhvbar}
\eeq
Likewise the more familiar Parke-Tarlor formula for MHV amplitude can be built up by adding positive particles.

Another example we like to consider is a particular BCFW diagram for a $n$-point amplitude with, say, $6$-point MHV on one side of the BCFW diagram, namely,
\beq \label{MHV6-pt}
A(\bar{1}2345|6 \cdots \hat{n})=A_{\rm MHV}(\bar{1},2,3,4,5,\hat{P}){1 \over P^2}A_R(-\hat{P},6,\cdots, \hat{n}),
\eeq
where we did not specify $A_R(-\hat{P},6,\cdots, \hat{n})$, in fact it can be anything, as we will discuss in section $4$. It is easy to check that this BCFW diagram is equivalent to the following ISL expression,
\beq
A(\bar{1}2345|6 \cdots \hat{n})=\big[ \mathcal{S}_+(345) \mathcal{S}_+(23'5')\mathcal{S}_+(12'5'')\mathcal{S}_-(n1'5''')\big]A_R(5'''', \cdots, n'),
\label{eg6NMHV}
\eeq
where $i'$ means it is shifted once according to the rules~(\ref{shift}) and (\ref{shift2}), $i''$ means shifted twice, and so on.

\section{Recursion relation for adding particles}

\subsection{MHV}

\begin{figure}[t]
\begin{center} 
\includegraphics[width=18cm]{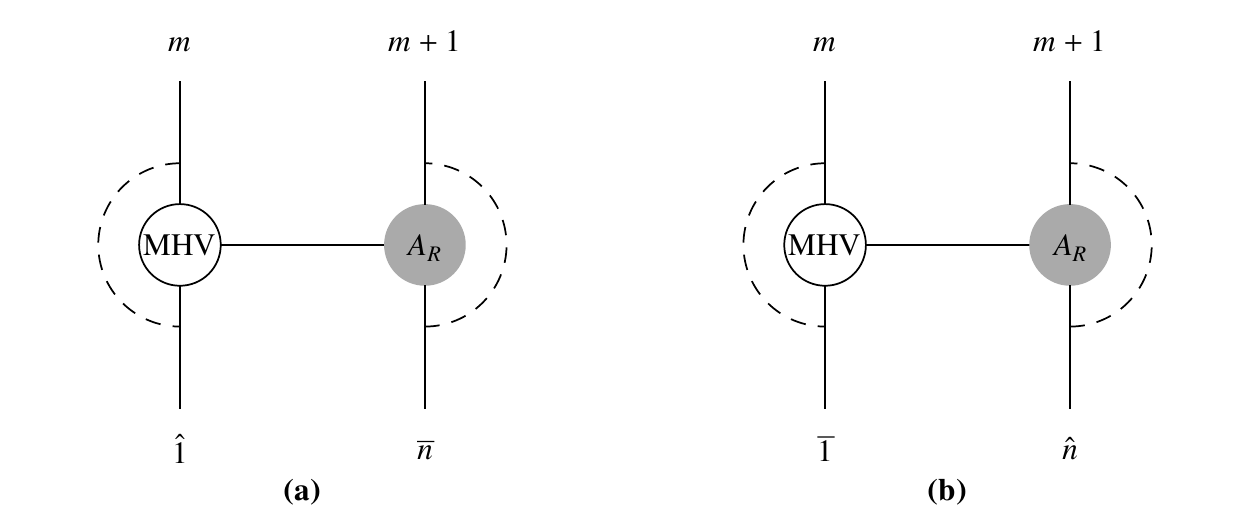}
\caption{\textbf{(a):} For the $\langle 1~n]$ shift we add particles $\{1, \ldots, m-1\}$ on the left side of the first diagram to make it $A_L^{{\rm MHV}}(\hat{1},\ldots, m, \hat{P})$ while the subamplitude $A_R$ on the right can be of any type. \newline
\textbf{(b):} For the $[1~n\rangle$ shift we add particles $\{1, \ldots, m-1\}$ on the left side of the first diagram to make it$A_L^{\rm MHV}(\bar{1},\ldots, m, \hat{P})$ .}
\label{MHVrecfig}\end{center}
 \end{figure}

We have seen a couple of examples of applying ISL to get amplitudes and BCFW diagrams, in this section we will present a systematic way of constructing a BCFW diagram by adding particles one at a time. Let us warm up with the simplest case when the BCFW diagram has a MHV amplitude on one side, see Fig.(\ref{MHVrecfig}). We will state the results first and will explain them shortly.

For $\langle 1~n ]$ BCFW shift, namely Fig.(\ref{MHVrecfig}.a) the way of adding the particles for this case is given as \beqa
\{ 1^+, 2^-, 3^+, \cdots, (m\!-\!1)^+ \},
\eeqa
the notation means that we add particle $1^+$ first, $2^-$ second, and so on until $(m\!-\!1)^+$. Just to simplify the notation, we define it as
\beqa \label{MHVhat}
 \hat{\A}^{(m)}_{\rm MHV} \equiv \{ 1^+, 2^-, 3^+, \cdots, (m\!-\!1)^+ \},
\label{AMHVshiftbar} 
 \eeqa
where $\A$ stands for ``Adding particles" and superscript $(m)$ denotes adding all possible particles labeled by $i$ such that $i < m$. We note here that in fact the ordering of the particles $3^+, \cdots, (m\!-\!1)^+ $ is not important, which is generally true that the ordering of the same helicity particles are not important. 

Similarly when we have $A_{\rm MHV}(\bar{1}, 2, \cdots, i, \hat{P})$ on one side of a BCFW diagram, namely for the $[ 1~n \rangle $ BCFW shift, see Fig.(\ref{MHVrecfig}.b), we add the particles as
 \beqa \label{MHVbar}
 \bar{\A}^{(m)}_{\rm MHV}\equiv \{ 1^-, 2^+, \cdots, (m\!-\!1)^+ \}.
 \label{AMHVshifthat}
 \eeqa
 
The above two statements can be understood as follows: let us first consider the $[1 ~ n\rangle$ shift, by construction we add the particle $1^-$ first, and by counting the ferminonic degrees for a MHV amplitude there must be one and only one negative particle, so the rest of the particles must be positive and hence we are led to \eqref{MHVbar}.

Similarly for the other case, $A_{\rm MHV}(\hat{1}, 2, \cdots, m, \hat{P})$, since the first particle now is $1^+$, then the next one added must be negative and the remaining should be all positive. We have proved the results of both cases, $\langle 1~n]$ shift and $[1 ~ n\rangle$ shift explicitly by comparing with BCFW recursion relations. 

In fact, the BCFW recursion relation in momentum-twistor\footnote{For a simple review on ISL in momentum-twistor space please see appendix.} is already in the ISL form for this simplest case we are considering. The tree-level BCFW recursion relations in momentum-twistor is given as\footnote{For more details about BCFW recursion relations in momentum-twistor and beyond tree-level, please see~\cite{ArkaniHamed:2010kv}. For comparison we have done reflection on the original formula, namely $a \rightarrow n-a+1$.} 
\be \label{BCFWMT}
&& M_{n,k}(1,\cdots,n) = M_{n-1,k}(2,\cdots,n) \cr
 &&~~~ +\! \sum_{n_{R},k_{R};j}[j\!+\!1~j~ 2 ~1~ n] M_{n_L,k_L}(\hat{1}_{j+1},\cdots,j,I_{j+1})M_{n_R,k_R}(I_{j+1},j\!+\!1,\cdots, n),
\ee
where $n_L+n_R=n+2, k_L+k_R=k-1,$ and the shifts are given as
\be
\hat{1}_{j+1} = (1 2) \bigcap (j j\!+\!1 n), I_{j+1} = (j j\!+\!1) \bigcap (n 1 2).
\ee
For the special case we are considering, namely when the amplitude on the left-hand-side is a MHV amplitude, we have $M_{n_L,k_L}(\hat{1}_{j+1},\cdots,j,I_{j+1})=1$, and Eq.~(\ref{BCFWMT}) reduces to 
\be \label{BCFWMTmhv}
M_{n,k}(1,\cdots,n) &=& M_{n-1,k}(2,\cdots,n) \cr
 &+& \sum_{n_{R},k_{R};j}[j\!+\!1~j~ 2 ~1~ n] M_{n_R,k_R}(I_{j+1},j\!+\!1,\cdots, n).
\ee
It is quite clear that the first term $M_{n-1,k}(2,\cdots,n)$ can be interpreted as adding a positive particle $1^+$, while the second term is corresponding to $\{ 1^+, 2^-, 3^+, \cdots, (j\!-\!1)^+ \},$ which is exactly the same as we described for $\langle 1~n]$ shift.

When $M_{n_L,k_L}(\hat{1}_{j+1},\cdots,j,I_{j+1})$ is beyond MHV, the BCFW recursion relations~(\ref{BCFWMT}) can not be so simply interpreted as ISL. Instead we will apply our results from MHV case to motivate the recursion relations for adding particles. It allows us to extend this program where our goal is to find a canonical configuration for adding particles to construct one side of the BCFW diagram when that is a general $(m+1)$ point amplitude. In the next couple of sections we motivate the systematic method of achieving our goal for the cases of next-maximally-helicity-violating (NMHV) and next-next-maximally-helicity-violating (NNMHV) amplitudes on one side of BCFW diagrams and finally in the subsequent section we give our general result for the N$^k$MHV case.

\subsection{NMHV}

\begin{figure}[t]
\begin{center} 
\hspace*{-2.0 cm}
\includegraphics[width=19cm]{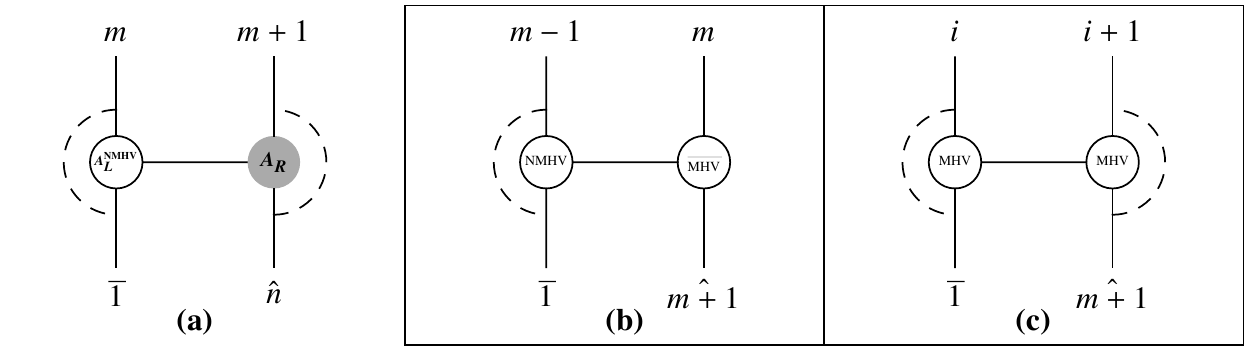}
\caption{\small{\textbf{(a):} For the $[1~n\rangle$ shift we add particles $\{1, \ldots, m-1\}$ on the left side of the this diagram to build up $A_L^{\rm NMHV}(\bar{1},\ldots, m, \hat{P})$ while $A_R$ on the right can be of any type.\newline
 \textbf{(b, c):} These are the corresponding BCFW contributions to the $(m+1)$ point subamplitude $A_L^{\rm NMHV}$ in {\bf (a)}.}} 
\label{NMHVfig1}\end{center}
 \end{figure}

In this section we consider the case when we have a NMHV amplitude on one side of a BCFW diagram. Let us start with the case when one side of the BCFW diagram is $A_{\rm NMHV}(\bar{1}, 2, \cdots, m, \hat{P})$, and we will denote $\bar{\A}^{(m)}_{\rm NMHV}$ as the way of adding particles for this case. To be a NMHV amplitude, $m$ must be greater than $3$. When $m=4$, one can easily check that the right way of adding particles is
\be \label{NMHV4pt}
\bar{\A}^{(4)}_{\rm NMHV} = \{ 1^-, 2^+, 3^- \}.
\ee
To understand the general case, let us first study the relevant $(m+1)$-point NMHV amplitude. The BCFW diagrams contributing to this amplitude are given in Fig.(\ref{NMHVfig1}).

Let us consider the two BCFW diagrams in the box separately. For the first BCFW diagram in the box, Fig.(\ref{NMHVfig1}.b), we have
\be 
A_{\rm NMHV}(\bar{1}, 2, \cdots, m-1, \hat{P}){1 \over P^2} A_{\overline{\rm MHV}}(-\hat{P}, m, \widehat{m\!+\!1}).
\ee
We note that the subamplitude $A_{\rm NMHV}(\bar{1}, 2, \cdots, m-1, \hat{P})$ can be viewed as built up by adding particles according to $\bar{\A}^{(m-1)}_{\rm NMHV}$. So the particles added at the last $(m\!-\!2)$ steps for $\bar{\A}^{(m)}_{\rm NMHV}$ are fully determined for this contribution, namely all particles appeared in $\bar{\A}^{(m-1)}_{\rm NMHV}$ except $1^-$. After those are determined, we are only left with the particles $(m-1)$ and $1$. By construction the particle $1^-$ must be added at the first step, and the particle $(m-1)$ must be positive. Putting all these together the analysis shows that if there is a ISL way of rewriting this BCFW diagram, the way of adding particles for this case is given as
\be 
\{1^-, (m-1)^+,\bar{\A}^{(m-1)}_{\rm NMHV}(\cancel {1^-})\},
\ee
where we use $\bar{\A}^{(m-1)}_{\rm NMHV}(\cancel {1^-})$ to denote all the particles appearing in $\bar{\A}^{(m-1)}_{\rm NMHV}$ except $1^-$.

Then let us consider the other contribution to this amplitude, Fig.(\ref{NMHVfig1}.c),
\be 
A_{\rm MHV}(\bar{1}, 2, \cdots, i, \hat{P}){1 \over P^2} A_{\rm MHV}(-\hat{P}, i+1, \cdots, \widehat{m+1}).
\ee
From previous section we know how to add particles when one side of BCFW diagram is a MHV amplitude. For the MHV amplitude on the left-hand-side, $A_{\rm MHV}(\bar{1}, 2, \cdots, i, \hat{P})$, the corresponding way of adding particles is given by $\bar{\A}^{(i)}_{\rm MHV}(\cancel{1^-})$; for $A_{\rm MHV}(-\hat{P}, i+1, \cdots, \widehat{m+1})$, it is given as $\{m^-, (i+2)^+, \cdots, (m-1)^+ \}$, however we should note that one cannot add particle $m$ in any step. So after $m^-$ is removed  this may be denoted as 
\be 
\mathcal{R}^{i-1}\left[ \hat{\A}^{(m-i)}_{\rm MHV}(\cancel{1^+,2^-}) \right],
\ee 
where $\mathcal{R}$ is a rotating operation, which does the cyclic shifting, $a \rightarrow a+1$ for any $a$ appeared in $\hat{\A}^{(m-i)}_{\rm MHV}(\cancel{1^+,2^-})$, and $\mathcal{R}^{i-1}$ mean we rotate the numbers $(i-1)$ times, i.e. $a \rightarrow a+i-1$. So from this analysis we learn in what order the particles should be added for the last $(m-3)$ steps and after this is done now we are left with only the particles $1$, $i$ and $(i+1)$. The order of their addition can be determined from the knowledge about the case of $m=4$, Eq.~(\ref{NMHV4pt}), which is simply $\{ 1^-, i^+, (i+1)^- \}$. 

In conclusion, if there is a ISL method of constructing this BCFW diagram, we find that the way of adding particles for this case must be given as
\be \label{NMHVbar1}
\{ 1^-, i^+, (i+1)^-, \mathcal{R}^{i-1}\left[ \hat{\A}^{(m-i)}_{\rm MHV}(\cancel {1^+, 2^-}) \right], \bar{\A}^{(i)}_{\rm MHV}(\cancel {1^-}) \}.
\ee
We can combine $\mathcal{R}^{i-1}\left[ \hat{\A}^{(m-i)}_{\rm MHV}(\cancel {1^+, 2^-}) \right]$ with $(i+1)^-$ and nicely arrive at $\mathcal{R}^{i-1}\left[ \hat{\A}^{(m-i+1)}_{\rm MHV}(\cancel {1^+}) \right]$. 
Putting all these together, we reach a recursion relation of adding particles for the case of having a NMHV subamplitude on one side of BCFW diagram
\beqa
\bar{\A}^{(m)}_{\rm NMHV} &=& \{1^-, (m-1)^+, \bar{\A}^{(m-1)}_{\rm NMHV}(1^-)\} \nn
&+& \sum^{m-2}_{i=2} \{ 1^-, i^+, \mathcal{R}^{i-1}\left[ \hat{\A}^{(m-i+1)}_{\rm MHV}(\cancel {1^+}) \right], \bar{A}^{(i)}_{\rm MHV}(\cancel {1^-}) \}.
 \label{NMHVrecbar}
\eeqa 
With the results from MHV case, it is straightforward to solve this recursion relation and find the general way of adding particles for this case which is given as
\beqa
 \bar{\A}^{(m)}_{\rm NMHV} &=&\sum^{m-1}_{i=4} \sum^{i-2}_{j=2} \{1^-, (m-1)^+, \cdots, i^+,j^+,(j+1)^-, (j+2)^+, \cdots,(i-1)^+  \} \cr 
&& + \sum^{m-2}_{i=2} \{ 1^-, i^+, (i+1)^-, (i+2)^+, \cdots,(m-1)^+, 2^+, \cdots,(i-1)^+  \}  \cr
&=& \sum^{m}_{i=4} \sum^{i-2}_{j=2} \{1^-, (m-1)^+, \cdots, i^+,j^+,(j+1)^-, R^+  \},
\label{NMHVrecbarsol1}
\eeqa
where we use $R^+$ to denote rest of the particles, namely particles except $\{1, (m-1), \cdots, i, j,(j+1) \}$, and they are all positive. There is no need to specify the ordering of these particles in $R^+$, since the ordering of adding the same helicity particles is not important.
 \begin{figure}[t]
\begin{center} 
\hspace*{-2.0 cm}
\includegraphics[width=19cm]{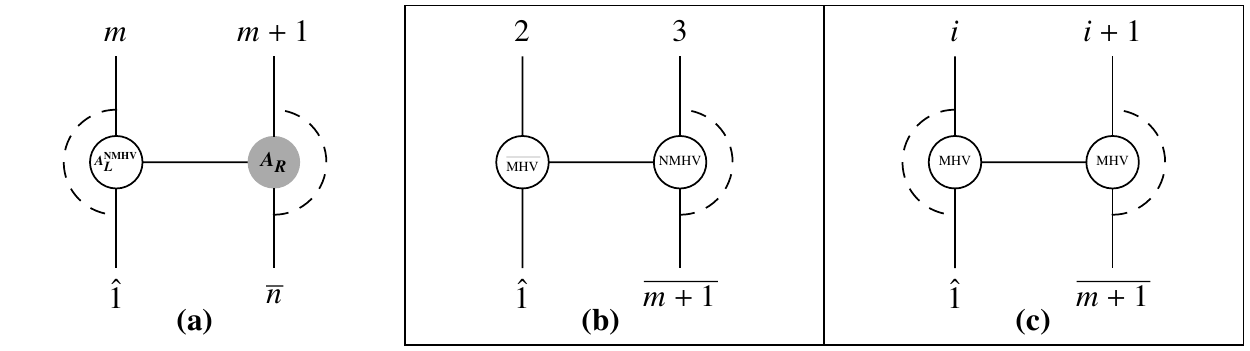}
\caption{\small{\textbf{(a):} For the $\langle1~n]$ shift we add particles $\{1, \ldots, m-1\}$ on the left side of the first diagram to build up $A_L^{\rm NMHV}(\hat{1},\ldots, m, \bar{P})$ while $A_R$ on the right can be of any type.\newline
 \textbf{(b, c):} These the corresponding BCFW contributions to the $(m+1)$ point subamplitude $A_L^{\rm NMHV}$ from {\bf (a)}.}} 
\label{NMHVfig2}\end{center}
 \end{figure}
 
Similarly we can motivate the recursion relations for the other kind of BCFW shift $\langle 1~n]$, namely the parity flipped version of the previous case, and we will denote it as $\hat{\A}^{(m)}_{\rm NMHV}$, see Fig.(\ref{NMHVfig2}). The BCFW diagrams relevant to this $(m\!+\!1)$-point NMHV amplitudes are given in Fig.(\ref{NMHVfig2}.b) and Fig.(\ref{NMHVfig2}.c). 

Again because of the knowledge of MHV amplitudes, the second case, Fig.(\ref{NMHVfig2}.c), leads to the following way of adding particles
\be 
\{1^+, i^-, \mathcal{R}^{i-1} \left[ \bar{\A}^{(m-i+1)}_{\rm MHV}(\cancel {1^-}) \right], \hat{\A}^{(i)}_{\rm MHV}(\cancel {1^+}) \}, 
\ee
or $\{1^+, i^-,(i+1)^+,(i+2)^+, \cdots,  (m-1)^+,  2^-, 3^+, \cdots, (i-1)^+ \}$. 

As for the contribution from the first diagram, Fig.(\ref{NMHVfig2}.b), let us look at some examples first. For the lowest case, when $m=5$, from the $5$-point NMHV amplitude appeared on the right-hand-side of Fig.(\ref{NMHVfig2}.b), we can easily determine that the last particle added should be $4^-$. After this one is fixed, we are left with a $m=4$ MHV situation, which should have $\{ 1^+, 2^-, 3^+ \}$, so we finally get 
\be 
\{ 1^+, 2^-, 3^+, 4^-  \}, 
\ee
and we note that the above formula can be nicely rewritten in a suggestive way as,
\be 
\{ 1^+, 2^-, \mathcal{R} \left[ \bar{\A}^{(4)}_{\rm NMHV}(\cancel {1^-}) \right] \}. 
\ee
Explicit calculations on higher-point cases show that this pattern preserves. So the result for this case is actually determined by $\bar{\A}^{(m-1)}_{\rm NMHV}$, and the way of adding particles can be simply summarized as 
\be 
\{ 1^+,2^-, \mathcal{R} \left[ \bar{\A}^{(m-1)}_{\rm NMHV}(\cancel {1^-}) \right] \}.
\ee
This allows us to write a recursion relation of adding particles for this case, which is given as
\beqa 
\hat{\A}^{(m)}_{\rm NMHV} &=& \{ 1^+,2^-, \mathcal{R} \left[ \bar{\A}^{(m-1)}_{\rm NMHV}(\cancel {1^-}) \right] \}  \nn
&+& \sum^{m-1}_{i=3} \{1^+, i^-, \mathcal{R}^{i-1} \left[ \bar{\A}^{(m-i+1)}_{\rm MHV}(\cancel {1^-}) \right], \hat{\A}^{(i)}_{\rm MHV}(\cancel {1^+}) \}.
\label{NMHVrechat}
\eeqa
It is also not difficult to solve the recursion relation, and we find the general way of adding particles for this case,
\be
\hat{\A}^{(m)}_{\rm NMHV} &=& \sum^{m-1}_{i=3}\{ 1^+, i^-, (i+1)^+, \cdots, (m-1)^+,2^-, R_1^+ \} \cr
&+& \sum^{m-1}_{i=4} \sum^{i-1}_{j=3} \{ 1^+, 2^-, (m-1)^+, \cdots, (i+1)^+,j^+, (j+1)^-, R_2^+ \},
\label{NMHVrechatsol1}
\ee
where $R^+_i$ is again used to denote particles left over in the corresponding curly brackets. 
\subsection{NNMHV}
 \begin{figure}[t]
\begin{center} 
\includegraphics[width=12cm]{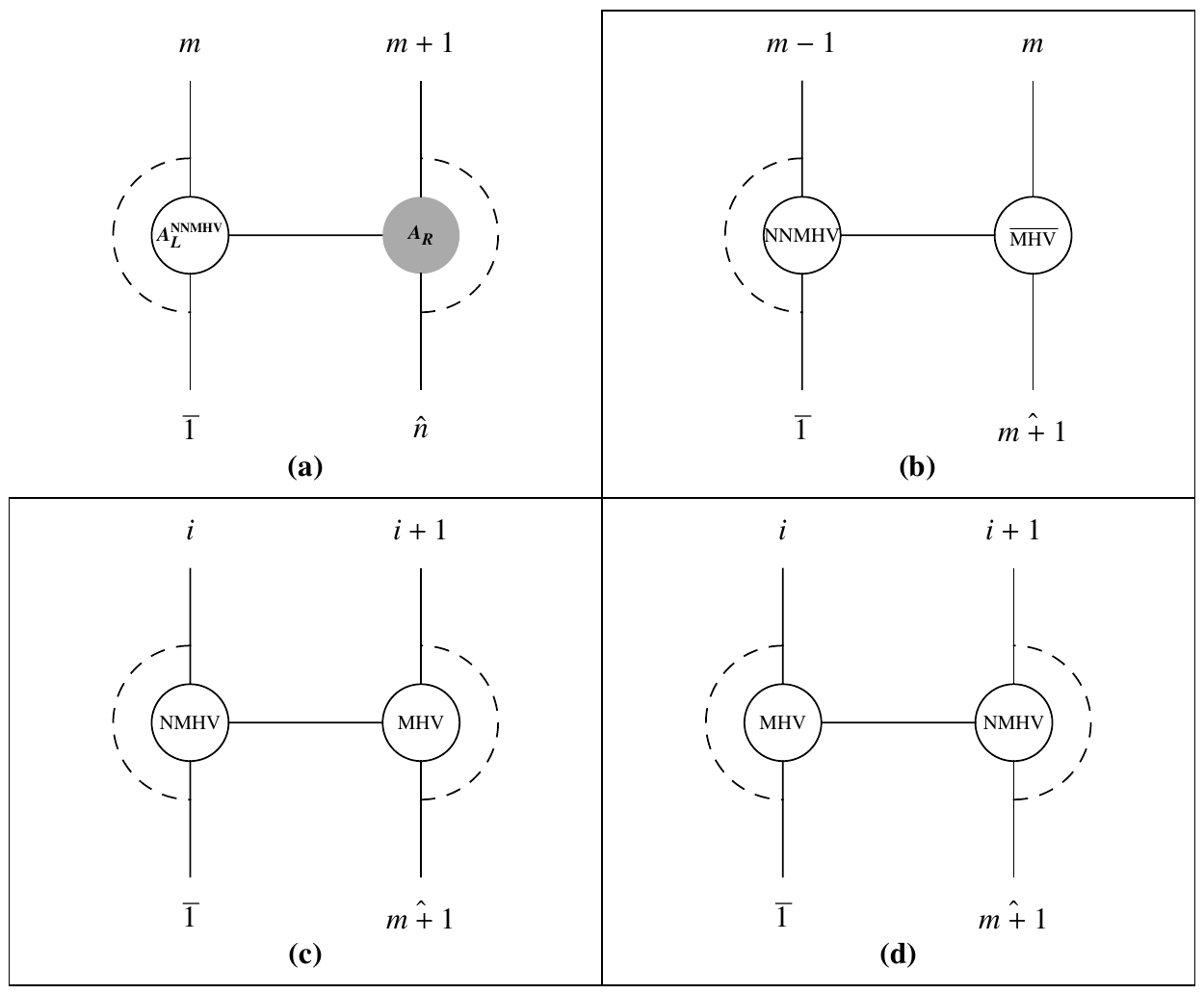}
\caption{\small{
\textbf{(a):} For the $[1~n\rangle$ shift we add particles $\{1, \ldots, m-1\}$ on the left side of the first diagram to make it $A_L^{\rm NNMHV}(\bar{1},\ldots, m, \hat{P})$ while the subamplitude $A_R$ on the right can be of any type.\newline
\textbf{(b, c, d):} In these three diagrams inside the box we consider the three different BCFW contributions that are possible for the $(m+1)$ point subamplitude $A_L^{\rm NNMHV} $ from {\bf (a)}.}} 
\label{N2MHVfig1}\end{center}
 \end{figure}

In this section we will study the case when we have a NNMHV subamplitude on one side of BCFW diagrams, as one more example before generalizing the recursion relations for a general N$^k$MHV case. To understand the ISL for this case, we need to study the corresponding NNMHV amplitude, which are given in the box of Fig.(\ref{N2MHVfig1}). We will use the knowledge from previous discussion to determine which particles should be added at certain last steps, consequently it motivates us to obtain the full recursion relations. Let us now study different contributions separately.

For the first BCFW diagram in the box Fig.(\ref{N2MHVfig1}.b), namely 
\be
A_{\rm NNMHV}(\bar{1}, 2, \cdots, (m\!-\!1), \hat{P}){1 \over P^2} A_{\overline {}\rm \overline{MHV}}(-\hat{P}, m, \widehat{m\!+\!1} ),
\ee
the same as NMHV case, it is quite straightforward to convince oneself that the particles should be added for this case is given as,
\be 
\{1^-, (m-1)^+, \bar{\A}^{(m\!-\!1)}_{\rm NNMHV}(\cancel {1^-}) \}. 
\ee
It is also not difficult to determine how the particles should be added for the second diagram Fig.(\ref{N2MHVfig1}.c), which is given as
\be
\{1^-, i^+, \mathcal{R}^{i-1} \left[ \hat{\A}^{(m-i+1)}_{\rm MHV}(\cancel {1^+})\right], \bar{\A}^{(i)}_{\rm NMHV}(\cancel {1^-}) \},
\ee
where the subamplitude, $A_{\rm NMHV}(\bar{1}, \cdots, \hat{P})$ in this BCFW diagram, contributes $\bar{\A}^{(i)}_{\rm NMHV}(\cancel {1^-})$. As in the case of Eq.~(\ref{NMHVbar1}) and (\ref{NMHVrecbar}), the contribution from $A_{\rm MHV}(-\hat{P}, \cdots, \widehat{m+1})$ can be combined with $(i+1)^-$, and finally leads to $\mathcal{R}^{i-1} \left[ \hat{\A}^{(m-i+1)}_{\rm MHV}(\cancel {1^+})\right]$ in the above equation.

Finally let us consider the contribution of the last diagram in the box Fig.(\ref{N2MHVfig1}.d). Let us study this case by starting from the simplest case when $m=5$, on the right-hand-side of this BCFW diagram, we have $A_{\rm NMHV}(-\hat{P}, 3,4,5,\hat{6})$. From $\hat{\A}^{(4)}_{\rm NMHV}$, we understand that this amplitude can be constructed by adding particles $\{ 6^+, 5^-, 4^- \}$, which means that the last step of ISL is to add $4^-$, as a consequence, if there is a ISL for this BCFW diagram, the particles added before $4^-$ should be $\{1^-, 2^+, 3^- \}$. In conclusion for this simplest case we find that the way of adding particles is given as,
\be
\{ 1^-, 2^+, 3^-, 4^- \},
\ee
which can be also be written as 
\be \label{4ptNNMHV}
\{ 1^-, 2^+, \mathcal{R} \left[ \hat{\A}^{(4)}_{\rm NMHV}(\cancel {1^+}) \right], \bar{\A}^{(2)}_{\rm MHV}(\cancel {1^-}) \},
\ee
where $\bar{\A}^{(2)}_{\rm MHV}(\cancel {1^-})$ is of course just empty.

Similar analysis and explicit calculations of higher-point cases show that this pattern, Eq.~(\ref{4ptNNMHV}), can be extended as a general result. As the above formula indicates, for the general case, the contribution from Fig.(\ref{N2MHVfig1}.d) leads to the following way of adding particles,
\be
\{ 1^-, i^+, \mathcal{R}^{i-1} \left[ \hat{\A}^{(m-i+1)}_{\rm NMHV}(\cancel {1^+}) \right], \bar{\A}^{(i)}_{\rm MHV}(\cancel {1^-}) \}.
\ee
So gathering all the informtion so far we arrive at a nice recursion relation for this case, which is given as
\be 
\bar{\A}^{(m)}_{\rm NNMHV} &=& \{1^-, (m-1)^+, \bar{\A}^{(m-1)}_{\rm NNMHV}(\cancel {1^-}) \} \\ \nonumber
&+& \sum^{m-2}_{i=4} \{1^-, i^+, \mathcal{R}^{i-1} \left[ \hat{\A}^{(m-i+1)}_{\rm MHV}(\cancel {1^+})\right], \bar{\A}^{(i)}_{\rm NMHV}(\cancel {1^-}) \} \\ \nonumber
&+& \sum^{m-3}_{i=2} \{ 1^-, i^+,  \mathcal{R}^{i-1} \left[ \hat{\A}^{(m-i+1)}_{\rm NMHV}(\cancel {1^+}) \right]  ,\bar{\A}^{(i)}_{\rm MHV}(\cancel {1^-})\}.
\label{N2MHVrecbar}
\ee

 \begin{figure}[t]
\begin{center} 
\includegraphics[width=13cm]{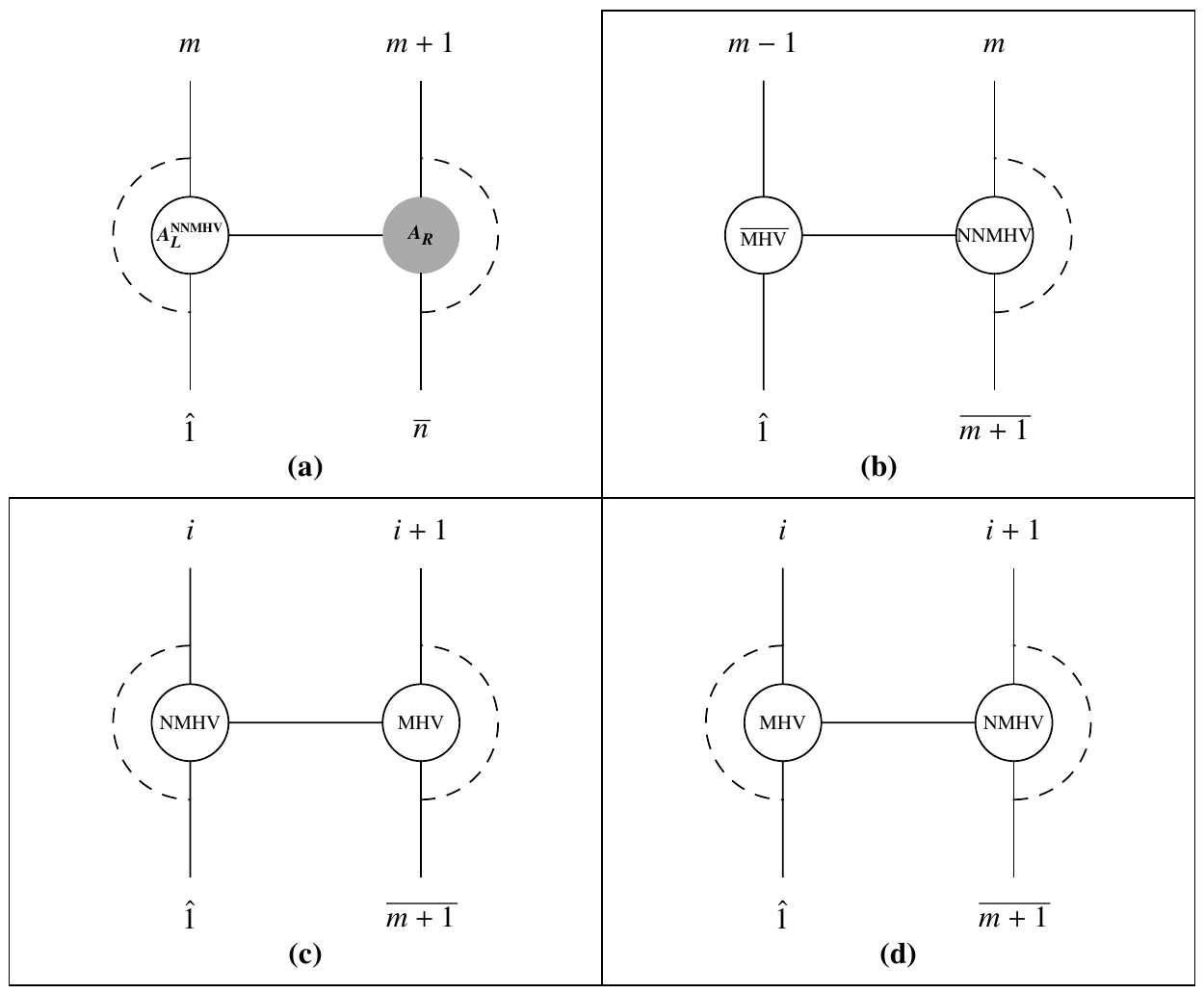}
\caption{\small{\textbf{(a):}For the $[1~n\rangle$ shift we add particles $\{1, \ldots, m-1\}$ on the left side of the first diagram to make it$A_L^{\rm NNMHV}(\bar{1},\ldots, m, \hat{P})$ while the subamplitude on the right can be of any type $A_R$.\newline
 \textbf{(b,c,d):} In these three diagrams inside the box we consider the three different BCFW contributions that are possible for the $(m+1)$ point subamplitude $A_L^{\rm NNMHV} $, from (a)}} 
\label{N2MHVfig2}\end{center}

 \end{figure}
Now let us concentrate on the parity inversion of above case, namely the $ \langle 1~n ]$ shift, see Fig.(\ref{N2MHVfig2}). It is straightforward to determine the last two type BCFW diagrams, Fig.(\ref{N2MHVfig2}.c) and Fig.(\ref{N2MHVfig2}.d), in the box of Fig.(\ref{N2MHVfig2}) by the same analysis as the case of $[1~n\rangle$ shift. The ways of adding particles determined by these two BCFW diagrams are given as the following sum,
\be
&& \sum^{m-1}_{i=4} \{1^+, i^-, \mathcal{R}^{i-1}\left[ \bar{\A}^{(m-i+1)}_{\rm MHV}(\cancel {1^-}) \right], \hat{\A}^{(i)}_{\rm NMHV}(\cancel {1^+}) \} \\ \nonumber 
&+& \sum^{m-3}_{i=3} \{1^+, i^-,\mathcal{R}^{i-1} \left[ \bar{\A}^{(m-i+1)}_{\rm NMHV}(\cancel {1^-}) \right], \hat{\A}^{(i)}_{\rm MHV}(\cancel {1^+}) \}.
\ee
While for the contribution from Fig.(\ref{N2MHVfig2}.b), after examining lots of non-trivial examples, we again observe, as in the case of NMHV amplitudes, that it is determined by lower-point $\bar{\A}^{(m-1)}_{\rm NNMHV}$ with an action of $\mathcal{R}$, namely
\be
\{ 1^+,2^-,\mathcal{R} \left[ \bar{\A}^{(m-1)}_{\rm NNMHV}(\cancel {1^-}) \right] \}. 
\ee
In summary that the final recursion relation of adding particles for this case is given as
\be 
\hat{\A}^{(m)}_{\rm NNMHV}&=& \{ 1^+,2^-,\mathcal{R} \left[ \bar{\A}^{(m-1)}_{\rm NNMHV}(\cancel {1^-}) \right] \}\\ \nonumber
 &+& \sum^{m-1}_{i=4} \{1^+, i^-, \mathcal{R}^{i-1}\left[ \bar{\A}^{(m-i+1)}_{\rm MHV}(\cancel {1^-}) \right], \hat{\A}^{(i)}_{\rm NMHV}(\cancel {1^+}) \} \\ \nonumber 
&+& \sum^{m-3}_{i=3} \{1^+, i^-,\mathcal{R}^{i-1} \left[ \bar{\A}^{(m-i+1)}_{\rm NMHV}(\cancel {1^-}) \right], \hat{\A}^{(i)}_{\rm MHV}(\cancel {1^+}) \}.
\label{N2MHVrechat}
\ee

\subsection{N$^k$MHV}
\begin{figure}[t]
\begin{center} 
\includegraphics[width=15cm]{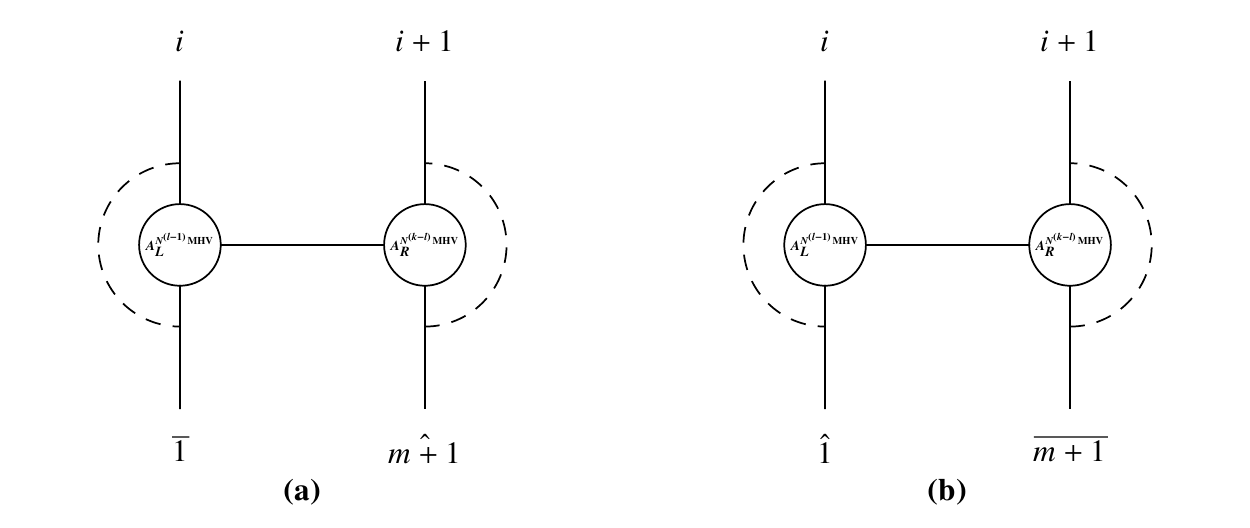}
\caption{\small{\textbf{(a):} Contribution to a $(m+1)$point N$^k$MHV amplitude for $[1~m+1\rangle$ shift.\newline 
\textbf{(b):}Contribution to a $(m+1)$point N$^k$MHV amplitude for $\langle 1~m+1]$ shift. }} 
\label{NkMHVfig1}\end{center}
 \end{figure}

In this section we will generalize the above recursion relations for th epreviously studied special cases to any kind of BCFW diagram of $\mathcal{N}=4$ super amplitudes. In this case we want to determine how to add particles when we have a BCFW diagram with $A_{{\rm N}^k{\rm MHV}}(\bar{1},2,\cdots, \hat{P})$ (and $A_{{\rm N}^k{\rm MHV}}(\hat{1},2,\cdots, \hat{P})$ for $\langle 1~n ]$ shift) on one side of the BCFW diagram. 

The way to determine the ISL for this case, namely the $[1~n \rangle$ shift, is to analyse the corresponding N$^k$MHV amplitude, which is given by Fig.(\ref{NkMHVfig1}.a). By considering $A_{{\rm N}^{l-1}{\rm MHV}}(\bar{1},2,\cdots,i, \hat{P})$ and $A_{{\rm N}^{k-l}{\rm MHV}}(-\hat{P},(i\!+\!1),\cdots, m, \widehat{m\!+\!1})$ separately and carrying out a similar analysis as the simpler cases of previous sections, it is not difficult to find that the way of adding particles for this typical BCFW diagram is given as,
\be
\{1^-, i^+, \mathcal{R}^{i-1} \left[ \hat{\A}^{(m-i+1)}_{{\rm N}^{l-1}{\rm MHV}}(\cancel {1^+})\right], \bar{\A}^{(i)}_{{\rm N}^{k-l}{\rm MHV}}(\cancel {1^-}) \},
\ee
which is a nice generalization of the special simpler examples we considered earlier. The final recursion relation is also straightforward to write down, which is given as
\beqa
\bar{\A}^{(m)}_{{\rm N}^{k}{\rm MHV}} &=& \{1^-, (m-1)^+, \bar{\A}^{(m-1)}_{{\rm N}^{k}{\rm MHV}}(\cancel {1^-}) \} \nn
&+& \sum_{l=1}^k \sum_{\substack{i=l+2\\(2\, {\rm for} \, l=1)}}^{m-k+l-2} \{1^-, i^+, \mathcal{R}^{i-1} \left[ \hat{\A}^{(m-i+1)}_{{\rm N}^{k-l}{\rm MHV}}(\cancel {1^+})\right], \bar{\A}^{(i)}_{{\rm N}^{l-1}{\rm MHV}}(\cancel {1^-}) \}. 
\label{recursionNkbar}
\eeqa
Similarly for the case of $\langle 1~n ] $ shift, Fig.(\ref{NkMHVfig1}.b), the recursion relation is given as,
\beqa
\hat{\A}^{(m)}_{{\rm N}^{k}{\rm MHV}} &=& \{1^+, 2^-, \mathcal{R} \left[ \bar{\A}^{(m-1)}_{{\rm N}^{k}{\rm MHV}}(\cancel {1^-}) \right] \}\nn
& +& \sum_{l=1}^k \!\!\! \sum_{i=l+2}^{\substack{(m-k+l-1 \, {\rm for} \, l=k)\\m-k+l-2}} \!\!\! \{1^+, i^-, \mathcal{R}^{i-1} \left[ \bar{\A}^{(m-i+1)}_{{\rm N}^{k-l}{\rm MHV}}(\cancel {1^-})\right], \hat{\A}^{(i)}_{{\rm N}^{l-1}{\rm MHV}}(\cancel {1^+}) \}. 
\label{recursionNkhat}
\eeqa
We would like to make a few comments on the above general recursion relations before we move on. We note that, as in the case of simpler examples of NMHV and NNMHV cases, the recursion relations are in fact coupled, namely $\bar{\A}^{(m)}$ and $\hat{\A}^{(m)}$ are determined recursively by each other. This fact is of course very natural since each BCFW diagram is made up of two subamplitudes (in left and right), which can be constructed by $\bar{\A}^{(m)}$ and $\hat{\A}^{(m)}$ separately. Although the pattern is quite intriguing, we were not able to prove this general recursion relation. However lots of non-trivial examples have been checked and we find that the amplitudes constructed from our recursion relations agree with those obtained by BCFW. In Appendix.B we present one such non-trivial examples and even more complicated cases had been worked out and matched with the BCFW results numerically. Further it is easy to convince oneself that $[1~ n \rangle$ shift and $\langle 1~ n ]$ shift should be related to each other by parity conjugation even though our recursion relations do not have manifest parity symmetry. This is indeed true and we find that
\beqa
\bar{\A}^{(m)}_{{\rm N}^{k}{\rm MHV}} =\mathcal{P} \big[ \hat{\A}^{(m)}_{{\rm N}^{m-k-3}{\rm MHV}} \big],
\eeqa
where $\mathcal{P}$ denotes the parity conjugation, namely it flips the signs of the particles $i^+ \leftrightarrow i^-$. This fact serves as a strong consistency check on our recursion relations. 

\subsection{Amplitudes from ISL}

We would like to conclude this section by summarising the prescription for constructing any BCFW diagram in $\N4$ SYM, for instance let us consider a typical BCFW term given as
\be 
A_L(\bar{1}, 2, \cdots, i,  \hat{P}){1 \over P^2} A_R(-\hat{P}, i\!+\!1, \cdots, \hat{n}).
\ee
 One can start with a three-point amplitude $A_{\overline{\rm MHV}}(i,i\!+\!1,n)$. By construction the first particle to be added is $1^-$, which is added between $i$ and $n$.\footnote{Or we could start with $A_{{\rm MHV}}(1, i,i\!+\!1)$, and the first particle to be added now is $n^+$, between $(i\!+\!1)$ and $1$.} We then keep adding particles between $1$ and $i$ according to the recursion relations of $\bar{\A}^{(i)}$ to fill $A_L(\bar{1}, 2, \cdots, i,  \hat{P})$ in the BCFW diagram, and separately $A_R(-\hat{P}, i\!+\!1, \cdots, \hat{n})$ can be filled by adding particles between $(i\!+\!1)$ and $n$ by applying the recursion relation of $\hat{\A}^{(n-i)}$ with the simple replacement of $k \rightarrow n-k+1$ for the elements in $\hat{\A}^{(n-i)}$.

 So in this way, any tree-level super amplitude in $\N4$ SYM theory can be schematically written in an ISL form, 
\be \label{ISLamp}
A_n =\sum_{i;L,R} (\prod_L \mathcal{S}'_L) (\prod_R \mathcal{S}'_R) A_{\overline{\rm MHV}}(i',i\!+\!1,n'),
\ee
where summation over $i$ is according to BCFW diagrammatic representation of the amplitudes, while products on $\mathcal{S}_L$ and $\mathcal{S}_R$ and summation on $L, R$ are determined by the recursion relation (\ref{recursionNkbar}) and (\ref{recursionNkhat}), finally $\mathcal{S}'$, $i'$ and $n'$ are used for the fact that the particles are shifted according to the rules of ISL. Here we like to stress that it is fairly easy to write down the actual amplitudes according to Eq.~(\ref{ISLamp}). In particular in the language of momentum-twistor, according to Eq.~(\ref{MT1}) adding a positive particle is fairly straightforward, in fact it does not change the form of the lower-point amplitude at all. For adding a negative particle we just need to multiply the lower-point amplitude with a R-invariant by Eq.~(\ref{MT2}), with some proper shifts on the corresponding particles, in Appendix B a non-trivial example is presented as we had also mentioned earlier.

\section{BCFW shifts from ISL}
From the previous sections we have seen that we can construct any BCFW diagram by adding particles according to ISL, where the canonical configuration for adding the required particles is given by the proposed recursion relations in \eqref{recursionNkbar} and \eqref{recursionNkhat}. Now let us stress the fact that once we have determined the canonical configuration to build a given BCFW diagram of our interest we use \eqref{shift} and \eqref{shift2} for shifting momenta and ferminonic coordinates at every step of adding a particle. For instance let us go back to the example we considered in section $2$, i.e. (\ref{MHV6-pt}) and we recall here that
\beq
A(\bar{1}2345|6, \cdots, \hat{n})=\big[ \mathcal{S}_+(345) \mathcal{S}_+(23'5')\mathcal{S}_+(12'5'')\mathcal{S}_-(n1'5''')\big]A_R(5'''', \cdots, n').
\eeq
If the BCFW and ISL form of $A(\bar{1},2,3,4,5|6, \cdots, \hat{n})$ have to match, as we had claimed, then the following equality between the BCFW and ISL quantities need to be satisfied, namely,
\beq
\hat{P}=5'''' ,\qquad \hat{n}=n',~~~{\rm and}\qquad \bar{1}=1'.
\eeq
Using \eqref{shift} and \eqref{shift2}, it can be easily shown that for this particular example the above equality holds. So here we see the very important fact that for adding particle from one side of a BCFW diagram, say for example here the left, the ISL shifts only affect those particles  in the right subamplitude $A_R$ which are adjacent to the $A_L$. In this case they are $5$ and $n$. So the ISL confguration obtianed for one side is blind to any configuration of particles on the other side and we will see this feature also being true for more general BCFW diagram. For those cases too, a similar equality between the BCFW and ISL expressions holds, as we will prove shortly. Let us first state the form of this equivalence. Let us consider the case where we can construct a BCFW diagram by ISL given in the following form,
\beq 
A_{L}(\bar{1}, 2, \cdots, i, \hat{P} ) {1 \over P^2} A_{R}(-\hat{P}, (i+1), \cdots, \hat{n}) \\ \nonumber
=\sum (\prod \mathcal{S}') A_{R}(i_s, (i+1), \cdots,(n-1), n_s),
\eeq
where the summation and products are determined by the proposed recursion relations, and here we use the subscript $s$ to denote the final shifted momenta obtained from ISL. The soft-factors are of course shifted too, which we denote as $\mathcal{S}'$.  The conclusion is that the following equalities between the ISL and BCFW shifted quantities hold,
\beqa
i_s = \hat{P}, \quad 1_s=\bar{1},\quad  n_s=\hat{n},
\eeqa
where the equality implies that both the bosonic momenta as well as the corresponding fermionic coordinate $\eta$'s satisfy the equality. 

To be more precise, let us consider the shift of type $[1~ n \rangle$, the result for the other kind of shift $\langle 1~ n ]$ can be obtained by parity conjugate. For this case, we first add $1^-$ to $ A_R(i, i\!+\!1, \cdots, n)$ and particles $i$ and $n$ are shifted as follows
\beqa \label{ISLshift1}
i &\rightarrow& i + {[1n] \over [in]} \lambda_1 \tilde{\lambda}_i=(1+i)+{s_{1i} \over \langle 1| i| n]} \lambda_1 \tilde{\lambda}_n ,\nn
n &\rightarrow& n + {s_{1i} \over \langle 1|i|n]} \lambda_1 \tilde{\lambda}_n,
\eeqa
where $s_{ij} \equiv (p_i + p_j)^2$ is the Mandelstam variable. From the recursion relation, \eqref{recursionNkbar} and \eqref{recursionNkhat}, we observe that whenever a negative particle, $j^-$, is added to a lower-point amplitude there is always a positive particle, $i^+$ already in front of it, being added at an earlier stage. When we say $i$ is in front of $j$ it is in the sense that these particles are cyclically ordered and hence $i<j$. This implies that $\lambda_1$ in above equation will not be shifted, since it can only be shifted by a negative particle, which is added next to $1$. The above conclusion precisely agrees with BCFW shifts for $[1~ n \rangle$ case, where only $\tilde{\lambda}_1$ and $\lambda_n$ are shifted. 

By the construction ISL preserves momentum conservation, so it is straightforward to see that Eq.~(\ref{ISLshift1}) will lead to the following equation when we finish adding all the particles,
\beqa
i_s &=&(1+2+\cdots +i)+{s_{12\cdots i} \over \langle 1| 2+\cdots +i| n]} \lambda_1 \tilde{\lambda}_n=\hat{P},\nn
n_s &=& n + {s_{12\cdots i} \over \langle 1| 2+\cdots +i| n]} \lambda_1 \tilde{\lambda}_n = \hat{n},
\eeqa
and we note that these are of course just the BCFW shifts for $\hat{P}$ and $\hat{n}$.

As for the fermionic coordinates, both $\eta_i$ and $\eta_n$ do not get shifted due to addition of the particle $1^-$, and $\eta_n$ will never be shifted by further addition of particles. So we see that there is no shift on ${\eta_n}$, which agrees also with the BCFW scenario.  

To determine the remaining shifted particles we can simply use the conservation laws, since the action of ISL keeps momenta and supercharge conserved. So by momentum conservation we get 
\be
1_s = -\left( 2+3+ \cdots + (i-1) \right) + i_s = 1 + {s_{12\cdots i} \over \langle 1| 2+\cdots +i| n]} \lambda_1 \tilde{\lambda}_n.
\label{1s}
\ee
The ISL shifts on ${\eta_i}_s$ can be similarly obtained by applying supercharge conservation,
\beq
{\lambda_1}_s {\eta_1}_s + \lambda_{2} \eta_{2} + \cdots + {\lambda_n}_s \eta_n = {\lambda_1} {\eta_1} + \lambda_{2} \eta_{2} + \cdots + {\lambda_n} \eta_n
\eeq
which gives us,
\beq
 {\eta_1}_s = \eta_1 - {\langle \hat{n} n\rangle \over \langle 1 n\rangle } \eta_n = \eta_1 - {s_{12\cdots i} \over \langle 1| 2+\cdots +i| n]} \eta_n.
 \label{etas}
\eeq
Both \eqref{1s} and \eqref{etas} agree with the results from BCFW shift of $[1~n\rangle$. 

We now summarize the conclusion from the above discussion. Here we observed that when we add the particles from one side (say $A_L$) of the BCFW diagram it affects only those two particles from the other side (say $A_R$) which are adjacent to this subdiagram ($A_L$). Moreover the effect of the successive ISL shifts on these two adjacent particles are exactly equivalent to the appropriate  BCFW shifts, which ensures that the amplitudes constructed by ISL agree with BCFW recursion relation. Furthermore, no other knowledge about the other side of the BCFW diagram is needed, which would turn out to be very important for the application of our discussion to form factors in the the following section.
 
\section{Constructing form factors}
\begin{figure}[t]
\begin{center} 
\includegraphics[width=15cm]{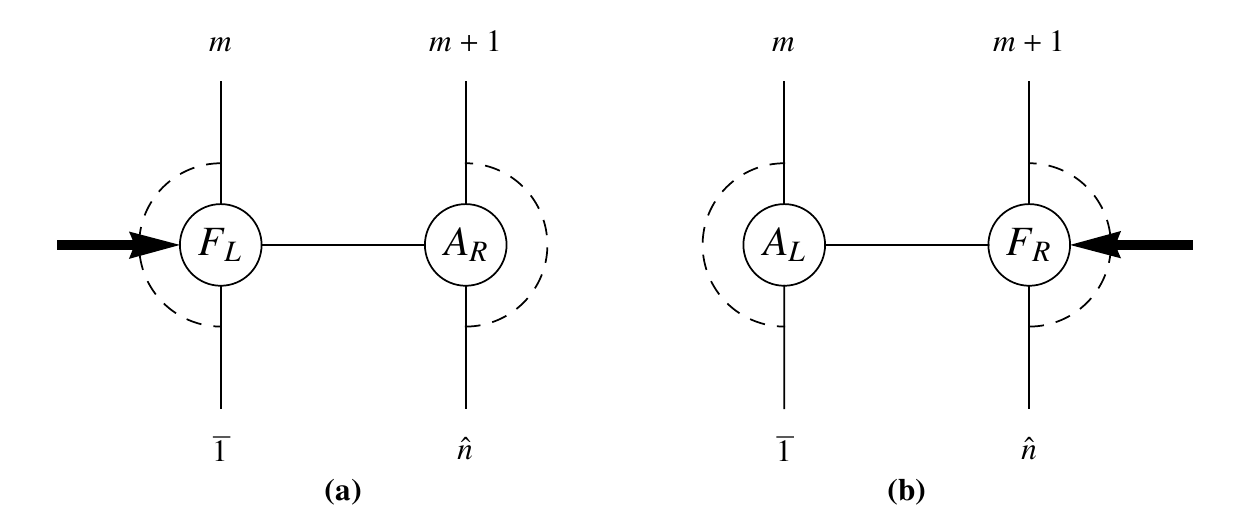}
\caption{\small{\textbf{(a,b):} The two possible BCFW diagrams for the $[1~n\rangle$ shift where $F$ is the form factor and $A$ is the amplitude.}} 
\label{FFbar}\end{center}
 \end{figure}

In this section, we will study another interesting object in $\mathcal{N}=4$ SYM, the form factors. We would like to apply the ISL we have developed for the amplitude to form factor as well. The object has been extensively studied in various aspects~\cite{Maldacena:2010kp,Brandhuber:2010ad,Brandhuber:2011tv,Bork:2011cj,Brandhuber:2012vm,Bork:2012tt}. Before going to the discussion of ISL for form factor, let us give a lightning review on form factors. We will closely follow the reference~\cite{Brandhuber:2011tv}.\footnote{For more details on form factors please see previously mentioned references.}

The form factors are the matrix elements of a gauge-invariant, composite operator between the
vacuum and some external scattering states,
\be
F(q; 1,2, \cdots, n) = \langle 1,2, \cdots, n| \mathcal{O}(q)|0 \rangle,
\ee
where $\langle 1,2, \cdots, n|$ are the external states, $|0 \rangle$ is the vacuum, and $\mathcal{O}(q)$ is a gauge invariant operator carrying momentum $q$; by momenta conservation we have
the sum of the momenta of external particles $\sum^n_{i=1} p_i = q $, and $q$ is not null, namely $q^2 \neq 0$. 
In $\mathcal{N}=4$ SYM, one can consider supersymmetric form factors by supersymmetrizing the external states as well as the operator. Here we can consider the full stress-tensor $\mathcal{T}(x, \theta^+, \theta^-)$ or we also allow  considering the chiral part of the stress-tensor $\mathcal{T}(x, \theta^+) = \mathcal{T}(x, \theta^+, \theta^-=0)$,\footnote{We have used harmonic superspace, and the form of $\mathcal{T}$ is not that important for our discussion, for the expression of $\mathcal{T}$ and details on harmonic superspace see for instance~\cite{Eden:2011yp, Eden:2011ku}} which we will do here. After Fourier transformation, the supersymmetrized form factor can be written as 
\be
F(q, \gamma^+ ; 1,2, \cdots, n) = \langle 1,2, \cdots, n| \mathcal{T}(q, \gamma^+)|0 \rangle,
\ee
where $\gamma^+$ is corresponding to the Fourier transformation of the fermionic valuable $\theta_+$. One can compute this object by various methods, including the well-known MHV rules and BCFW recursion relation. The supersymmetric BCFW recursion relations for form factor is simply given as
\be \label{BCFWFF}
F(q, \gamma^+; 1,2,\cdots, n) &=&\sum_i \big[ \int d^4 \eta F(q, \gamma^+ ; \bar{1}, 2, \cdots, m, \hat{P})A(-\hat{P}, (m\!+\!1), \cdots, \hat{n}) \\ \nonumber
&+& \int d^4 \eta A(\bar{1}, 2, \cdots, m, \hat{P})F(q, \gamma^+ ;-\hat{P}, (m\!+\!1), \cdots, \hat{n}) \big] ,
\ee
with the same usual supersymmetric BCFW shifts, see (\ref{bcfwshift}) and (\ref{Pbcfwshift}). The BCFW diagram is given in Fig.(\ref{FFbar}).

For the case of MHV, it is straightforward to find the solution that form factor for this simple case is given as,
\be 
F_{\rm MHV}(q, \gamma^+ ; 1,2, \cdots, n) =
 { \delta^4(\sum^n_{i=1} \lambda_i \tilde{\lambda}_i - q) \delta^4(\sum^n_{i=1} \lambda_i \eta^-_i) \delta^4(\sum^n_{i=1} \lambda_i \eta^+_i - \gamma^+)  \over \langle 1 2 \rangle \langle 23 \rangle \cdots \langle n 1 \rangle }.
\ee
As one can note that except the conservation delta-functions, the above formula is exactly the same as the famous Parke-Taylor formula for the scattering amplitudes. And indeed form factors resemble many properties of amplitudes, in particular one important property which is relevant to our discussion is that form factors have exactly the same soft limit by taking external on-shell particles to be soft as the amplitudes do.\footnote{One can also take the operator $\mathcal{O}$ to be soft, see the discussion in~\cite{Bork:2011cj,Brandhuber:2011tv}} 

We observe that one side of the BCFW recursion of form factors is always given by an amplitude as shown in Eq.~(\ref{BCFWFF}). As we discussed in previous section that if we add particles from one-side of a BCFW diagram, it is immaterial what is the type of object on the other side of the BCFW diagram, so it is quite clear that form factors can also be fully constructed by ISL. The idea of ISL we described in previous sections for the scattering amplitudes can apply to form factors directly without any essential modification. The way of adding particles for form factors is precisely the same as in the case of scattering amplitudes, except that now we need to add particles from both sides of BCFW diagrams, which is not a problem at all because we have derived recursion relations for adding particles with two type of BCFW shifts, namely $\bar{1}$ and $\hat{1}$. So for adding particles from left-hand-side of BCFW diagrams of the form factor, it is exactly the same as the amplitudes. But now we also have to add particles from the other side with the following simple replacement rule, $i \rightarrow n+1-i$.                                                                                                                          From the above discussion, we find that schematically the BCFW recursion relation of form factors, Eq.~(\ref{BCFWFF}) can be written in a ISL form, 
\be \label{ISLFF}
F(q, \gamma^+;1,2,\cdots, n) &=& \sum_{m;L,R} \big[ (\prod_R \mathcal{S}'_R) F(q, \gamma^+; 1', 2, \cdots, m, (m\!+\!1)') \\ \nonumber
&+& (\prod_L \mathcal{S}'_L) F(q, \gamma^+; m', (m+1), \cdots,n\!-\!1,n') \big].
\ee
Let us consider a simple example to illustrate the above formula~(\ref{ISLFF}). As case of amplitudes, we consider maximally non-MHV (MNMHV) form factor, which was considered in~\cite{Dixon:2004za,Brandhuber:2011tv}. It is a form factor with the self-dual field strength ${\rm Tr}(F^2_{\rm SD})$
and all negative helicity gluons states,
\be \label{anti-MHVFF}
F_{\rm MNMHV}(q;1,\cdots, n) = \langle 1 \cdots n|{\rm Tr}(F^2_{\rm SD})|0 \rangle |_{\rm MNMHV},
\ee
and the result of this special form factor is given as
\be
F_{\rm MNMHV}(q;1,\cdots, n) = \delta^4(\sum^n_{i=1} \lambda_i \tilde{\lambda}_i - q) {q^4 \over [12] \cdots [n1]}\eta^4_1 \cdots \eta^4_n.
\ee
There is only a two-particle channel BCFW diagram for this case, so it is easy to see that it can be written as an ISL form, 
\be \label{anti-MHVFFISL}
F_{\rm MNMHV}(q;1,2,\cdots, n) = \mathcal{S}_-(n~1~2)F_{\rm MNMHV}(q;2',3,\cdots, n').
\ee
It is easy to check that the formula (\ref{anti-MHVFF}) indeed satisfies the above recursion relation, Eq.~(\ref{anti-MHVFFISL}). Alternatively one can start with a two-point MNMHV (it is just MHV for this special case) form factor and then keep adding negative particles to arrive at (\ref{anti-MHVFF}). 

\section{Conclusion}
In this paper we had shown that any tree-level superamplitudes as well as supersymmetric form factors in $\N4$ SYM can be constructed by ISL. With guidance from BCFW recursion relations and detailed study of nontrivial examples, we are able to obtain a set of recursion relations, which give us the configuration for adding particles  in order to construct any BCFW diagram in $\N4$ SYM. Consequently, these recursion relations allow us to generate any tree-level superamplitudes and form factors by ISL method. It is a fascinating insight that for $\N4$ SYM theory, the restrictions imposed  due to soft limit is sufficient to determine the full scattering amplitudes, at least at tree-level. We note that scattering amplitudes constructed by ISL make both Yangian symmetry and soft-limit manifest. The application of ISL method to form factors indicates that there may also exist hidden symmetry in form factors, given that similarity has been noticed between form factors and the scattering amplitudes in $\N4$ SYM.

So convincingly the ISL method provides a new way of uncovering the deep mathematical structure of scattering amplitudes in $\N4$ SYM. The idea of ISL has been a extremely useful tool for constructing Grassmannian formalism for $\N4$ SYM, while in this paper we provide another intriguing use of it. And the picture we developed, of adding particles to lower-point amplitudes for generating higher-point amplitudes, seems to be intrinsically geometrical, which may be closely related to the Polytopes picture for the scattering amplitudes~\cite{ArkaniHamed:2010gg}. Moreover, as we had shown in the paper, ISL method may also be relevant as another useful tool for carrying out efficient computations in $\mathcal{N}=4$ SYM theory, both for the scattering amplitudes and form factors.

It would be of great interest to extend this idea for the scattering amplitudes to other theories and also beyond tree-level and leading singularities. Some interesting attempt has been made for $\mathcal{N}=8$ super gravity, where most of the progress so far had been for the simplest MHV case. The authors of~\cite{BoucherVeronneau:2011nm} were able to write the fist non-MHV amplitude, six-point NMHV gravity amplitude in a ISL form, however the form seems quite complicated to be amenable for further generalization to higher-point cases. One important difficulty for applying ISL method to gravity amplitudes is that there is no color-ordering. One naive guess would be to apply the ISL to the ordered subamplitudes, where the BCFW recursion relations for ordered subamplitudes have the same structure as those of Yang-Mills amplitudes~\cite{Drummond:2009ge},  but we leave these interesting questions to be addressed in future.


\begin{acknowledgments}

We are grateful to  Jacob Bourjaily, Mathew Bullimore, Laurentiu Rodina, Antun Skanata, Gabriele Travaglini, Anastasia Volovich and Gang Yang for very useful conversations. We would especially like to thank Anastasia Volovich for collaborations during earlier stages of this project. C.W. would like to thank Brown University, HET Group for the hospitality where part of the work was done. This work is supported in part by
by the US Department of Energy under contracts
DE-FG02-91ER40688 (Task A) and DE-FG02-11ER41742 (Early Career Award), 
the US National Science Foundation under grant
PHY-0643150 (PECASE) and Sloan Research Fellowship.
C.W. would like to acknowledge the support of the STFC Standard Grant ST/J000469/1 ``String
Theory, Gauge Theory and Duality".

\end{acknowledgments}
 \appendix

 \section{ISL in Momentum Twistor}

It is often more convenient to consider ISL in momentum twistor. In~\cite{Nima-allloop} the authors provide a general prescription for constructing an $n$-point Yangian invariant, $Y_{n,k}$ by adding a particle to the $n-1$ point Yangian invariant. We will give a brief review of their ideas here.\footnote{See \cite{Nima-allloop,Bourjaily:2010wh} for more details.} Building Yangian invariants can be done in two ways, either it is $k$ preserving operation as in 
\be \label{MT1}
Y'_{n,k}(\mt(1),\ldots, \mt(n-1), \mt(n)) =Y_{n-1,k}(\mt(1),\ldots \mt(n-1)),
\ee 
or both $k$ and $n$ increasing operation as in 
\be \label{MT2}
Y'_{n,k}(\ldots,\mt(n-1), \mt(n), \mt(1), \ldots) =[n\!-\!2~ n\!-\!1 ~ n~ 1~ 2]Y_{n-1,k-1}(\ldots ,\widehat{\mt(n-1)}, \widehat{\mt(1)},\ldots).
\ee 
We can see that the first case is pretty straightforward as it does not change the functional form of the Yangian invariants. For the second type, the lower point invariant have their super momentum twistors adjacent to the added particle, i.e. the $n_{th}$ particle, deformed by the following shifts,
\beqa
\widehat{\mt(1)}&=& \mt(1)\ab{2\,\,n-2\,\,n-1\,\,n}+\mathcal{Z}_2\ab{n-2\,\,n-1\,\,n\,\,1}; \nn
\widehat{\mt(n-1)} &=& \mathcal{Z}_{n-2}\ab{n-1\,\,n\,\,1\,\,2}+\mathcal{Z}_{n-1}\ab{n\,\,1\,\,2\,\,n-2},
\label{MTshifts}
\eeqa
 and we have the R-invariant defined as
 \beqa
[a\,\ b\,\ c\,\ d \,\ e]=\frac{\Gd^{0|4}(\eta_a \ab {b\,\ c\,\ d\,\ e } + \rm{cyclic})}{\ab {a\,\ b\,\ c\,\ d }\ab {b\,\ c\,\ d\,\ e }\ab {c\,\ d\,\ e\,\ a }\ab {d\,\ e\,\ a\,\ b }\ab {e\,\ a\,\ b\,\ c }}.
\label{Rinv} 
 \eeqa
Translate to the language of usual spinor formalism the first case~(\ref{MT1}) is corresponding to adding a positive particle in section $2$, while the second one~(\ref{MT2}) is corresponding to adding a negative particle. 

\section{Example of the ISL recursion relations}
\begin{figure}[t]
\begin{center} 
\includegraphics[width=6cm]{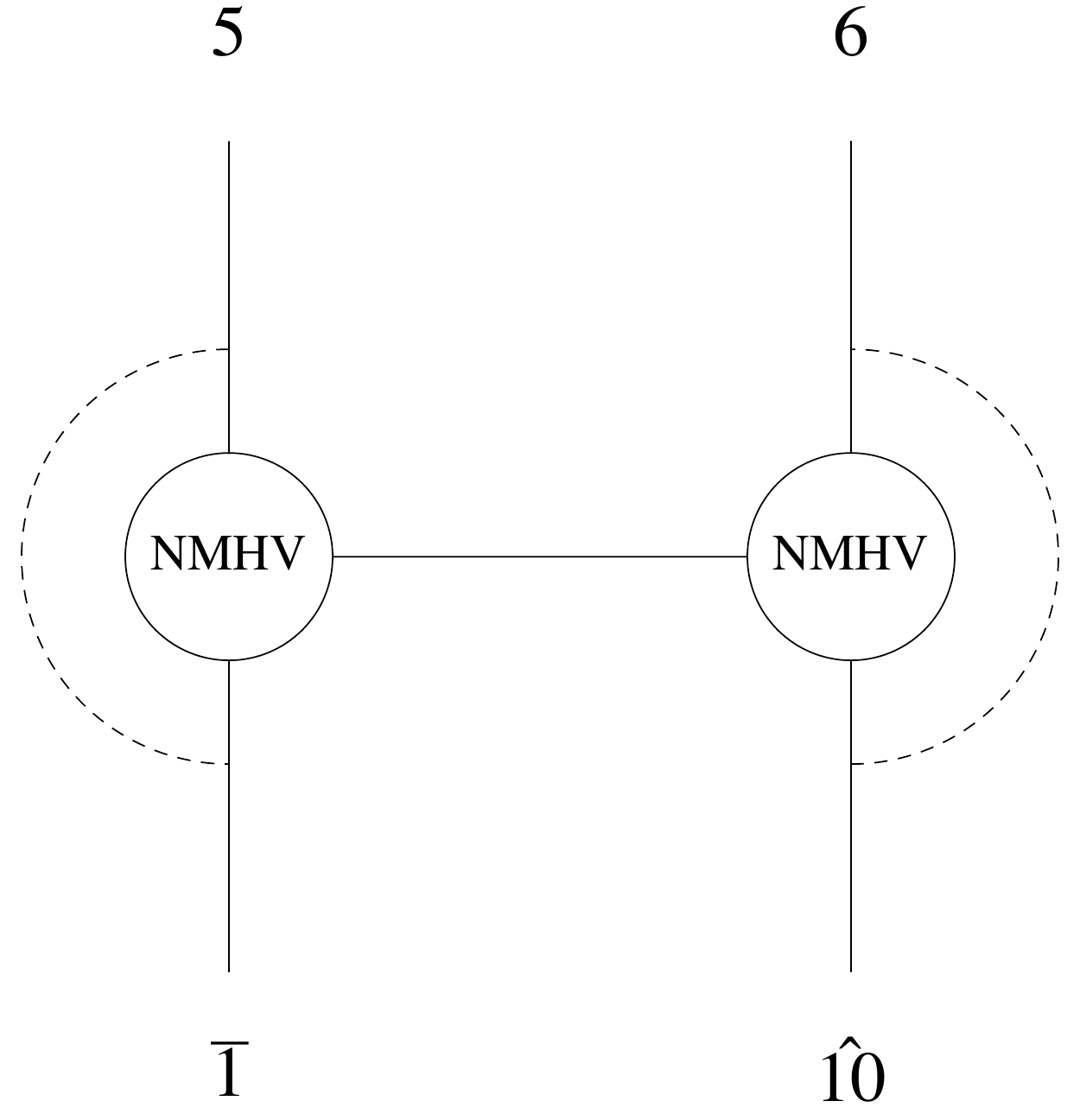}
\caption{\small{A particular BCFW diagram occuring in $10$ point N$^3$MHV amplitude}} 
\label{NMHVfigexample}\end{center}
 \end{figure}
Here we consider one concrete example of how a BCFW diagram can be built up from three-point amplitude by the recursion relation. The BCFW diagram is of the form 
\be 
A_{{\rm NMHV}}(\bar{1},2,3,4,5, \hat{P}) \times A_{{\rm NMHV}}(-\hat{P},6,7,8,9,\hat{10}),
\ee
and we will study this in the language of momentum-twistor, which is more compact. According to our prescription~(\ref{ISLamp}) we will start with $A_{\overline{\rm MHV}}(5,6,10)$ and by adding $1^-$, we get $ A_{\rm MHV}(1,5,6,10) = 1$ in the language of momentum-twistor. According to the recursion relations, we can gradually add particles between $1$ and $5$, as well as $6$ and $10$, the final results are listed below,
\beqa
\hspace{-1.0 cm}
\begin{array}{l}
 \left\{1^-,4^+,2^+,3^-\right\}_L\left\{9^-,8^+,7^-\right\}_R\equiv  [9~ 8~ \textbf{7}~ 6~ 5][1~ 10~ \textbf{9}~ 6~ 5] [1~2~\textbf{3}~4~5]\\
\left\{1^-,2^+,3^-,4^+\right\}_L\left\{9^-,8^+,7^-\right\}_R\equiv  [9~ 8~ \textbf{7}~ 6~ 5][1~ 10~ \textbf{9}~ 6~ 5] [1~2~\textbf{3}~5~\hat{\hat{6}}] \\
\left\{1^-,3^+,4^-,2^+\right\}_L\left\{9^-,8^+,7^-\right\}_R\equiv  [9~ 8~ \textbf{7}~ 6~ 5][1~ 10~ \textbf{9}~ 6~ 5] [1~3~\textbf{4}~5~\hat{\hat{6}}]\\
\left\{1^-,4^+,2^+,3^-\right\}_L\left\{9^-,8^+,7^-\right\}_R\equiv [1~ 10~ \textbf{9}~ 8~ 7][1~ \hat{10}~ \hat{\textbf{8}}~ 6~ 5] [1~2~\textbf{3}~4~5]\\
\left\{1^-,2^+,3^-,4^+\right\}_L\left\{8^-,7^+,9^-\right\}_R\equiv  [1~ 10~ \textbf{9}~ 8~ 7][1~ \hat{10}~ \hat{\textbf{8}}~ 6~ 5][1~2~\textbf{3}~5~\hat{\hat{6}}]\\
\left\{1^-,3^+,4^-,2^+\right\}_L\left\{7^-,9^-,8^+\right\}_R\equiv  [1~ 10~ \textbf{9}~ 8~ 7][1~ \hat{10}~ \hat{\textbf{8}}~ 6~ 5] [1~3~\textbf{4}~5~\hat{\hat{6}}]\\
\left\{1^-,4^+,2^+,3^-\right\}_L\left\{9^-,8^+,7^-\right\}_R\equiv[1~ 10~ \textbf{9}~ 7~ 6][1~ \hat{10}~ \hat{\textbf{7}}~ 6~ 5] [1~2~\textbf{3}~4~5]\\
\left\{1^-,2^+,3^-,4^+\right\}_L\left\{8^-,7^+,9^-\right\}_R\equiv[1~ 10~ \textbf{9}~ 7~ 6][1~ \hat{10}~ \hat{\textbf{7}}~ 6~ 5] [1~2~\textbf{3}~5~\hat{\hat{6}}] \\
\left\{1^-,3^+,4^-,2^+\right\}_L\left\{7^-,9^-,8^+\right\}_R\equiv [1~ 10~ \textbf{9}~ 7~ 6][1~ \hat{10}~ \hat{\textbf{7}}~ 6~ 5] [1~3~\textbf{4}~5~\hat{\hat{6}}],
\end{array}
\eeqa
where the left-hand-side denotes the way of adding particles (both from left and right of the BCFW diagram) to the three-point amplitude $A_{\overline{\rm MHV}}(5,6,10)$, while the right-hand-side means the answer in terms of R-invariant, and hats denote the shifts according to Eq.~(\ref{MTshifts}). We have also checked much more complicated examples.

 \section{Two-particle channel BCFW and ISL in gravity}

One can also generalize the discussion in section $2$ for Yang-Mills amplitudes to the gravity amplitudes~\cite{BoucherVeronneau:2011nm}. In gravity the three-point $\overline{\rm MHV}$ amplitude is given as
\be
M_L(1,2,\hat{P}) = {\delta^8 (\eta_1 [2 \hat{P}] + \eta_2 [\hat{P} 1] - \eta_{\hat{P}}  [12])  \over [12]^2 [2 \hat{P}]^2 [\hat{P} 1]^2},
\ee
and the corresponding soft factor is defined as following,
\be \label{softfactor}
\mathcal{G}_+ (n~1~2) \equiv  M_L(1,2,\hat{P}) {1 \over s_{12}} = {\langle n 2 \rangle^2 [2 1] \over \langle n 1 \rangle^2 \langle 1 2 \rangle} = \mathcal{S}^2_+(n~ 1~ 2)s_{12},
\ee
with the same shifts on particles $2$ and $n$ as Eq. (\ref{shift}).

Since there is a bonus relation between gravity amplitudes\cite{ArkaniHamed:2008gz,Spradlin:2008bu,He:2010ab}, the above soft factor can be further simplified. For instance for a MHV amplitude, under the shift Eq.~(\ref{bcfwshift}), we have BCFW recursion relation and the bonus relation,
\be
M_2+M_3+\dots + M_{n-1}=M,   \\  \nonumber
z_2 M_2+z_3 M_3+ \dots + z_{n-1} M_{n-1} =0,
\ee
which allows us to remove $M_{n-1}$ in the whole amplitude $M$ and get an extra bonus factor~\cite{Spradlin:2008bu}
\be
B_{n-1}=1-{\langle 1 i \rangle  \langle n n-1 \rangle \over \langle n i \rangle \langle 1 n-1 \rangle}={\langle 1 n \rangle  \langle i n-1 \rangle \over \langle n i \rangle \langle 1 n-1 \rangle},
\ee
multiply this bonus factor with soft factor $\mathcal{G} (n,1,i)$ in (\ref{softfactor}), we arrive at more familiar result
\be
\mathcal{G}_B (n~1~i)={\langle n i \rangle  \langle i n-1 \rangle [i 1]\over \langle n 1 \rangle \langle 1 i \rangle  \langle 1 n-1 \rangle },
\ee
which is the soft factor used in~\cite{Bern:1998sv,Nguyen:2009jk}, and the corresponding ISL recursion relation from this soft factor is the same as the one appeared in~\cite{Hodges:2011wm}, which was originally obtained from $\mathcal{N}=7$ BCFW.~\cite{Elvang:2011fx} Similar consideration can be done for negative graviton.

\end{document}